\documentclass[useAMS,usegraphicx,usenatbib]{mn2e}
\usepackage{aas_macros}
\usepackage[a4paper,centering, totalwidth=520pt, totalheight=700pt]{geometry}
\usepackage[fleqn]{amsmath} 
\usepackage{graphicx}
\usepackage{amssymb}
\usepackage{amsmath}
\usepackage{enumerate}
\usepackage{color}
\usepackage{hhline}

\usepackage{fixltx2e}[1999/12/01]

\usepackage[hyperindex]{hyperref}
\hypersetup{
breaklinks = {false},
colorlinks = {false},
linkcolor={black},
pdfpagemode = {None}, % prevents the automatic announcement of bookmarks
pdfborder = {0 0 1},
pdftitle = {Subhalo Abundance Matching and Assembly bias in EAGLE},
pdfsubject = {SHAM},
pdfauthor = {chaves-montero,angulo},
pdfkeywords = {large-scale structure of the Universe -- dark matter -- galaxies: haloes -- galaxies: formation -- galaxies: evolution}
}

\newcommand{\kpc}{\ensuremath{\mathrm{kpc}}}
\newcommand{\Mpc}{\ensuremath{\mathrm{Mpc}}}

\newcommand{\Mass}{\ensuremath{ M_{\odot}}}

\newcommand{\Mstar}{ \ensuremath{M_{\rm star}} }
\newcommand{\Mgas}{ \ensuremath{M_{\rm gas}} }
\newcommand{\Mhalo}{ \ensuremath{M_{\rm halo}} }
\newcommand{\Msun}{ \ensuremath{\rm M_{\odot}} }
\newcommand{\probmv}{\ensuremath{P(\log_{10} {\rm M}_{\rm star}|\log_{10} V_i)}}

\newcommand{\Vmax}{ \ensuremath{ V_{\rm max}} }
\newcommand{\Vpeak}{ \ensuremath{ V_{\rm peak}} }
\newcommand{\Vinfall}{ \ensuremath{ V_{\rm infall}} }
\newcommand{\Vrelax}{ \ensuremath{ V_{\rm relax}} }
\newcommand{\Vcirc}{ \ensuremath{ V_{\rm circ}} }

\newcommand{\kms}{\ensuremath{\mathrm{km\;s}^{-1}}}

\title[Testing SHAM with EAGLE]
{Subhalo abundance matching and assembly bias in the EAGLE simulation}
\begin{document}

\author[J. Chaves-Montero et al.]{
\parbox[h]{\textwidth} {Jon\'as Chaves-Montero$^{1}$, Raul E. Angulo$^{1}$, 
Joop Schaye$^{2}$, Matthieu Schaller$^{3}$, Robert A. Crain$^{4}$, Michelle Furlong$^{3}$ \& Tom Theuns$^{3}$.}
\vspace*{6pt}
  \\$^{1}$Centro de Estudios de F\'isica del Cosmos de Arag\'on, Plaza San Juan 1,  Planta-2, 44001, Teruel, Spain.
  \\$^{2}$ Leiden Observatory, Leiden University, P.O. Box 9513, 2300 RA Leiden, the Netherlands.
  \\$^3$ Institute for Computational Cosmology, Dep.~of Physics, Univ.~of Durham, South Road, Durham, DH1 3LE, UK.
  \\$^{4}$ Astrophysics Research Institute, Liverpool John Moores University, 146 Brownlow Hill, Liverpool, L3 5RF.
}

\maketitle

\date{\today}

\begin{abstract}
Subhalo abundance matching (SHAM) is a widely-used method to connect galaxies
with dark matter structures in numerical simulations. SHAM predictions agree
remarkably well with observations, yet they still lack strong theoretical
support. We examine the performance, implementation, and assumptions of SHAM
using the EAGLE project simulations. We find that $\Vrelax$, the highest value
of the circular velocity attained by a subhalo while it satisfies a relaxation
criterion, is the subhalo property that correlates most strongly with galaxy
stellar mass ($\Mstar$). Using this parameter in SHAM, we retrieve the
real-space clustering of EAGLE to within our statistical uncertainties on scales
greater than $2\;\Mpc$ for galaxies with $8.77<\log_{10}(\Mstar[\Msun])<10.77$.
Conversely, clustering is overestimated by $30\%$ on scales below $2\;\Mpc$ for
galaxies with $8.77<\log_{10}(\Mstar[\Msun])<9.77$ because SHAM slightly
overpredicts the fraction of satellites in massive haloes compared to EAGLE. The
agreement is even better in redshift-space, where the clustering is recovered to
within our statistical uncertainties for all masses and separations.
Additionally, we analyse the dependence of galaxy clustering on properties other
than halo mass, i.e. the assembly bias. We demonstrate assembly bias alters the
clustering in EAGLE by $20\;\%$ and $\Vrelax$ captures its effect to within
$15\;\%$. We trace small differences in the clustering to the failure of SHAM as
typically implemented, i.e. the $\Mstar$ assigned to a subhalo does not depend
on i) its host halo mass, ii) whether it is a central or a satellite. In EAGLE
we find that these assumptions are not completely satisfied.
\end{abstract}


\section{Introduction}

The clustering of galaxies offers an excellent window to explore galaxy
formation processes and the fundamental properties of our Universe. On small
scales, correlation functions can inform us about the way in which galaxies
populate dark matter (DM) haloes and thus about the efficiency of star
formation and the importance of environmental effects. On large scales, the
clustering of galaxies can be used to constrain cosmological parameters and the
law of gravity. On even larger scales, the observed distribution of galaxies is
sensitive to the physics of inflation and relativistic effects. By using
correlation functions of different orders and at distinct scales, degeneracies
among several parameters can be broken, providing even tighter constrains on
all the aforementioned quantities.

To extract the information encoded in the clustering of galaxies, we need
accurate predictions for a given cosmological scenario and galaxy formation
model. However, obtaining the correct galaxy distribution is a difficult task,
especially at small scales where besides highly non-linear dynamics,
gravitational collapse, mergers, dynamical friction, and tidal stripping;
baryonic processes such as star formation, feedback, and ram pressure are at
play. Consequently, one needs to resort to numerical simulations to obtain
accurate predictions for galaxy clustering \citep[see][for a
review]{2012PDU.....1...50K}. 

Two types of approach can be followed. The first is to simulate the joint
evolution of DM and baryons by solving the Poisson and Euler equations coupled
with recipes for unresolved physical processes (e.g. star and black hole
formation). Although this approach currently yields the most direct predictions
for the distribution of galaxies, it is computationally infeasible to simulate
large cosmological volumes with adequate resolution for calculating accurately
the galaxy clustering on scales on the order of $100\;h^{-1}\,\Mpc$. In
addition, simulations have only recently begun to produce populations of
realistic galaxies \citep{2014MNRAS.444.1518V,2015MNRAS.446..521S}. 

The second approach is to simulate only gravitational interactions and to
predict the galaxy clustering a posteriori. This is justified by leading
theories of galaxy formation, where DM plays the dominant role in determining
the places where galaxies form and merge. Gravity-only simulations (a.k.a.
DM-only simulations) are computationally less expensive and can thus follow
sufficiently large volumes to enable the correct interpretation of observational
surveys. This is an important advantage since, for instance, to model galaxy
clustering on scales beyond $100\;h^{-1}\,\Mpc$, it is necessary to perform
$N$-body simulations of volumes in excess of $1\;h^{-3}\,\rm{Gpc}^3$
\citep{Angulo08}. The disadvantage is that the predictions for galaxy clustering
are more uncertain because the relation between galaxies and DM haloes is not
straightforward.

{\it Subhalo abundance matching} \citep[SHAM,
e.g.,][]{2004MNRAS.353..189V,2006ApJ...643...14S, 2006ApJ...647..201C} is a
widely-used method to populate gravity-only simulations with galaxies. The
original version of SHAM assumes an injective and monotonic relation between
galaxies and self-bound DM structures based on a set of specified properties.
SHAM usually links galaxies to DM structures using stellar mass as galaxy
property and a measure of subhalo mass, such as circular velocity, as subhalo
property. More recent implementations introduce stochasticity into the relation
to make the model more realistic \citep[e.g.,][]{2010ApJ...717..379B,
2011ApJ...742...16T, 2013ApJ...771...30R, 2014MNRAS.443.3044Z}. Then, SHAM
places each galaxy at the centre-of-potential of its corresponding subhalo and
assumes that each galaxy has the same velocity as the centre-of-mass of its
linked subhalo. SHAM thus makes predictions for the clustering of galaxies, but
not for any physical properties such as stellar mass, star formation rate,
metallicity, etc.

SHAM predictions have been shown to agree remarkably well with observations
\citep[e.g.][]{2006ApJ...647..201C, 2010MNRAS.404.1111G, 2010MNRAS.403.1072W,
2010ApJ...710..903M, 2010ApJ...717..379B, 2011ApJ...742...16T,
2012ApJ...754...90W, 2013MNRAS.432..743N, 2013ApJ...771...30R}.  For instance,
\cite{2006ApJ...647..201C} showed that SHAM reproduces the observed galaxy
clustering over a broad redshift interval ($0<z<5$). More recently,
\cite{2013ApJ...771...30R} achieved a simultaneous fit to the clustering and
the conditional stellar mass function measured in SDSS.
\cite{2013MNRAS.436.1142S} even used this model to constrain cosmological
parameters, finding values in good agreement with those obtained from more
established methods.

Despite these successes, the comparison with simulations of galaxy formation has
not been so encouraging. \cite{2008ApJ...678....6W} found that the galaxy
clustering predicted by SHAM only agrees with that of a hydrodynamical
simulation beyond $1\;h^{-1}\,\Mpc$. On smaller scales, the differences were of
the order of a few. \cite{2012MNRAS.423.3458S} extended the previous study using
two hydrodynamic simulations with different feedback models. They found that the
clustering predicted by SHAM exceeded that of their most realistic simulation by
more than a factor of 2 on scales below $0.5\;h^{-1}\,\Mpc$. Finally, in a
direct comparison with two semi-analytic models of galaxy formation,
\cite{2015arXiv150206614C} found that SHAM performs well at some galaxy number
densities, but not at others.

It is therefore not clear whether SHAM is able to match the observed galaxy
clustering because it makes accurate assumptions (i.e. the physical relation
between subhaloes and galaxies) or because some implementations employ free
parameters (e.g. a scatter between subhalo and galaxy properties or a cut-off
in the fraction of satellite galaxies) that provide enough freedom to become
insensitive to them. The importance of the information being decoded, added to
the fact that the amount and accuracy of clustering data will increase
dramatically over the next decade due to the emergence of wide-field galaxy
surveys (e.g. DES, HETDEX, eBOSS, JPAS, DESI, EUCLID, and LSST), makes it
crucial to critically test the assumptions underlying SHAM.

In this paper we will employ the state-of-the-art hydrodynamical simulations
EAGLE \citep{2015MNRAS.446..521S,2015MNRAS.450.1937C} to study the SHAM
technique in detail. Our objectives are threefold, i) to seek the most accurate
implementation of SHAM, ii) to directly test the underlying assumptions, and
iii) to assert how accurately SHAM can predict galaxy clustering.

We will propose $\Vrelax$, defined as the maximum of the circular velocity of a
DM structure along its entire history while it fulfils a relaxation criterion,
as the best subhalo property with which to perform SHAM. We will show that this
definition captures the best qualities of previously proposed implementations
while mitigating their disadvantages and reducing the number of problematic
cases. As a consequence, $\Vrelax$ shows the strongest correlation with the
simulated stellar mass of EAGLE galaxies. 

We will show that SHAM is able to reproduce the clustering properties of stellar
mass selected galaxies in the EAGLE simulation (which successfully reproduces
many properties of observed low-$z$ galaxies). For the stellar mass range
investigated ($10^{8.77}<\Mstar[\Msun]<10^{10.77}$), the agreement is better
than $10\;\%$ on scales greater than $2\;\Mpc$, and better than $30\;\%$ on
smaller scales. The agreement is particularly good for massive galaxies and in
redshift space, for which we do not find statistically significant difference
between the clustering predicted by SHAM and EAGLE. This is remarkable given
that we explore almost two orders of magnitude in spatial scale and four in
clustering amplitude.

Additionally, we will pay attention to the so-called ``assembly bias'': the
dependence of the clustering of DM haloes on properties other than mass
\citep{2005MNRAS.363L..66G,2006ApJ...639L...5Z,2006ApJ...652...71W,
2007MNRAS.374.1303C, 2007MNRAS.377L...5G,2008ApJ...686...41Z,
2008ApJ...687...12D,
2008MNRAS.389.1419L,2011MNRAS.412.1283L,2012MNRAS.426L..26L,
2014MNRAS.443.3044Z,2014MNRAS.443.3107L,2014arXiv1404.6524H}. We will show that
assembly bias is present in both EAGLE and SHAM galaxies, increasing the
clustering amplitude by $20\;\%$ on scales from 2 to $11\;\Mpc$. To our
knowledge, this is the first detection of assembly bias in a hydrodynamical
simulation. This result supports the idea that {\it Halo Occupation Distribution
models} \citep[HOD, e.g.,][]{2000MNRAS.318..203S, 2000MNRAS.318.1144P,
2001ApJ...546...20S}, which are a phenomenological parametrization for the
number of galaxies hosted by haloes of a given mass, introduce bias in the
calculation of galaxy clustering when they assume that halo occupation is a
function only of halo mass.

Finally, we will track the small residual differences in the clustering of SHAM
and EAGLE galaxies to the failure of a key assumption of SHAM (as
commonly implemented): for the same $\Vrelax$, central and satellite subhaloes
host the same galaxies independently of their host halo mass. We will find that
this supposition is broken due to the influence of the environment and the star
formation that satellite galaxies experience after having been accreted. Both
effects correlate with the mass of the DM host, which suggests that future SHAM
implementations that employ both host halo mass and $\Vrelax$ could yield even
more accurate predictions for the clustering signal.

Our paper is organized as follows. In \S2 we describe the simulations, halo and
galaxy catalogues, and merger trees that we use. In \S3 we discuss different
implementations of SHAM and introduce $\Vrelax$, a new proxy for stellar mass.
In \S4 we analyse the accuracy with which SHAM can predict the galaxy satellite
fraction, host halo mass, clustering, and assembly bias.  We discuss the
limitations of SHAM in \S5. We conclude and summarize our most important
results in \S6.

\section{Numerical Simulations}

In this section we provide details of the main datasets that we employ. 
This includes a brief description of the numerical simulations, halo and 
galaxy catalogues, merger trees, and of a technique to identify the same
structures in our hydrodynamical and gravity-only simulations.

\begin{table}
\begin{center}
\caption{EAGLE/DMO cosmological and numerical parameters. The cosmological
parameter values are taken from \protect\cite{2014A&A...571A...1P,2014A&A...571A..16P}.}
\label{tabular:params}
\begin{tabular}{c|c}
\hline \hline
Parameter&EAGLE/DMO\\ \hline 
$\Omega_{\rm m}$&0.307\\ 
$\Omega_\Lambda$&0.693\\ 
$\Omega_{\rm b}$&0.04825\\ 
$H_0 [\kms\text{ Mpc}^{-1}]$&67.77\\ 
$\sigma_8$&0.8288\\ 
$n_{\rm s}$&0.9611\\
Max. proper softening [\kpc]&0.70\\
Num. of baryonic particles&$1504^3/-$\\
Num. of DM particles&$1504^3/1504^3$\\
Initial baryonic particle mass $[10^7 \Msun]$&$0.181/-$\\
DM particle mass $[10^7 \Msun]$&$0.970/1.150$\\ \hline
\end{tabular}
\end{center}
\textbf{Notes.} $\Omega_{\rm m}$, $\Omega_\Lambda$, and $\Omega_{\rm b}$ are the
densities of matter, dark energy, and baryonic matter in units of the critical
density at redshift zero. $H_0$ is the present day Hubble expansion rate,
$\sigma_8$ is the linear fluctuation amplitude at $8\;h^{-1}\,\Mpc$, and $n_{\rm
s}$ is the scalar spectral index.  
\end{table}

\subsection{The EAGLE suite}

The simulations we analyse in this paper belong to the ``Evolution and Assembly
of Galaxies and their Environment'' project
\citep[EAGLE;][]{2015MNRAS.446..521S,2015MNRAS.450.1937C} conducted by the
Virgo consortium. EAGLE is a suite of high-resolution hydrodynamical
simulations aimed at understanding the formation of galaxies in a cosmological
volume. The runs employed a pressure-entropy variant
\citep{2013MNRAS.428.2840H} of the Tree-PM smoothed particle hydrodynamics code
{\small GADGET3} \citep{2005MNRAS.364.1105S}, the time step limiters of
\cite{2012MNRAS.419..465D}, and implement state-of-the-art subgrid physics
\citep[as described by][]{2015MNRAS.446..521S}, including metal-dependent
radiative cooling and photo-heating \citep{2009MNRAS.393...99W}, chemodynamics
\citep{2009MNRAS.399..574W}, gas accretion onto supermassive black holes
\citep{2013arXiv1312.0598R}, star formation
\citep{2008MNRAS.383.1210S}, stellar feedback \citep{2012MNRAS.426..140D}, and
AGN feedback. 

The EAGLE suite includes runs with different physical prescriptions,
resolutions, and volumes. Here, we study the largest simulation, which follows
$1504^3$ gas particles and the same number of DM particles inside a periodic
box with a side length of $100 \text{ Mpc}$. The large volume and high
resolution of this simulation are essential for a careful analysis of SHAM. The
cosmological parameters used in EAGLE are those preferred by the analysis of
{\it Planck} data (Table \ref{tabular:params}).  This implies a gas particle mass
equal to $1.81 \times 10^6 \Msun$ and a DM particle mass equal to $9.70 \times
10^6 \Msun$. We highlight that EAGLE is well suited to this study because it was 
calibrated to reproduce the galaxy stellar mass function at $z\sim0$. The agreement
with observations is especially good over the mass range that we will analyse here
\citep[fig. 4 of][]{2015MNRAS.446..521S}.

The $100 \text{ Mpc}$ box was resimulated including only gravitational
interactions and sampling the density field with $1504^3$ particles of mass
$1.15 \times 10^7 \Msun$. Hereafter, we refer to this simulation and its
hydrodynamical counterpart as DMO and EAGLE, respectively. The cosmological and
some of the numerical parameters employed in these simulations are provided in
Table \ref{tabular:params}. 

\subsection{Catalogues and mergers trees}

In each simulation, haloes were identified using only DM particles and a
standard friends-of-friends ({\small FOF}) group-finder with a linking
parameter $b=0.2$ \citep{1985ApJ...292..371D}. Gas and star particles are
assigned to the same FoF halo as their closest DM particle.  For each FoF halo
we compute a spherical-overdensity mass, $M_{200}$, defined as the mass inside
a sphere with mean density equal to 200 times the critical density of the
Universe, $\rho_{\rm crit}(z)$;

\begin{equation}
M_{200}=\frac{4\pi}{3}200 \rho_{\rm crit} r_{200}^3,
\end{equation}

\noindent where $r_{200}$ is the radius of the halo, $\rho_{\rm
crit}(z)=\frac{3\,H^2(z)}{8\pi G}$, $G$ is the gravitational constant, and
$H(z)$ is the value of the Hubble parameter $H(z)=H_0 \sqrt{\Omega_{\rm m}
(1+z)^3 + \Omega_\Lambda}$. 

Self-bound structures inside FoF haloes, termed subhaloes, were identified
using all particle types and the {\small SUBFIND} algorithm
\citep{2001MNRAS.328..726S, 2009MNRAS.399..497D}.  Hereafter, we will refer to
the subhalo located at the potential minimum of a given FoF halo as the
``central'', to any other structures as ``satellites'', and to subhaloes with
more than one star particle as EAGLE ``galaxies''.

The position of each galaxy is assumed to be that of the particle situated at
the minimum of the gravitational potential of the respective subhalo. The galaxy
velocity is assumed to be that of the centre of mass of the subhalo \footnote{We
checked that the mean difference between the bulk velocity of DM particles and
star particles in the inner 30 $\kpc$ for the subhaloes with $8.77<\Mstar
[\Msun]<10.77$ is smaller than 10 $\kms$.}. The stellar mass, $\Mstar$, is the
total mass of all star particles linked to a given EAGLE galaxy. The gas mass
($\Mgas$) and the dark matter mass ($M_{\rm DM}$) are computed in the same
manner but using gas particles or dark matter particles, respectively. We
verified that our results are insensitive to the exact definition of $\Mstar$:
we repeated our analysis defining $\Mstar$ as the mass inside a sphere of 20,
30, 40, 50, 70, or 100 $\kpc$ radius. We found that different mass definitions
only produces sub-percent differences in the galaxy clustering.

We employ ``merger trees'' to follow the evolution of haloes and subhaloes,
their mass growth, tidal stripping, mergers, as well as transient effects in
their properties. Our trees were built using the algorithm described in
\cite{2014MNRAS.440.2115J}, employing $201$ snapshots for DMO and $29$
snapshots for EAGLE. In both simulations the output times were approximately
equally spaced in $\log(a)$ for $a>0.2$, where $a$ is the cosmic scale factor.

Finally, we note that to avoid problems related to subhalo fragmentation and
spurious structures, we remove from our analysis satellites without resolved
progenitors.
 
\subsection{EAGLE and DMO crossmatch}
\label{sec:crossmatch}

EAGLE and DMO share the same initial conditions, so we expect
roughly the same non-linear objects to form in both simulations. This is a
powerful feature: it enables us to identify the EAGLE galaxy that a given DMO
subhalo is expected to host, and thus, to probe directly the assumptions of
SHAM.

In practice, we link DMO subhaloes to EAGLE galaxies following the process
described by \cite{Schaller2015}; see also \cite{2014MNRAS.442.2641V}. For every
subhalo in EAGLE we select the 50 most-bound DM particles. If we find a subhalo
in DMO which shares at least half of them, the link is made. We confirm the link
if, repeating the same process starting from each DMO subhalo, we identify the
same pair. We only search the pairs with more than 174 DM particles in each
simulation, which corresponds to a minimum halo mass of $2\times10^9\Msun$ in
DMO. This procedure yields a catalogue of 13687 galaxies with
$10^{8.77}<\Mstar[\Msun]<10^{10.77}$.

In Table \ref{table:cross} we list the fraction of successfully matched
centrals and satellites, for four stellar mass bins. Overall, the match is
successful for more than $90\;\%$ of centrals in EAGLE, independently of their
mass. The success rate drops to $68-80\;\%$ for satellites, with low-mass
satellites showing the lowest percentage. This is a consequence of the finite
mass resolution of the simulations (see also Appendix A), the mass loss due to
interactions with the host halo, small differences in the timing at which
mergers happen, and the high-density environment in which they reside. 

\begin{table}
\begin{center}
\caption{Number of central and satellite EAGLE galaxies for four stellar mass
bins. In parentheses we provide the percentage of EAGLE galaxies with a
counterpart in DMO.} 
\label{table:cross}
\begin{tabular}{@{\extracolsep{2pt}}c|r|r|@{}}
\hline\hline
\vspace{2pt}
$\log_{10}(\Mstar[M_\odot])$&\multicolumn{2}{c}{EAGLE}\\ \cline{2-3} 
\vspace{-7pt}\\
 &\multicolumn{1}{c}{Central}&\multicolumn{1}{c}{Satellites}\\ \hline
$8.77-9.27$   & 3954 ($92\;\%$) & 3475 ($68\;\%$) \\
$9.27-9.77$   & 2550 ($92\;\%$) & 2068 ($74\;\%$) \\
$9.77-10.27$  & 1551 ($94\;\%$) & 1247 ($76\;\%$) \\
$10.27-10.77$ & 968  ($92\;\%$) & 652  ($80\;\%$) \\ \hline
\label{tabular:bins}
\end{tabular}
\end{center}
\end{table}

\section{Subhalo abundance matching}
\label{sec:SHAM}

In this section we discuss different SHAM flavours and their implementation in DMO.

\begin{figure*}
\begin{center}
\includegraphics[width=0.475\textwidth]{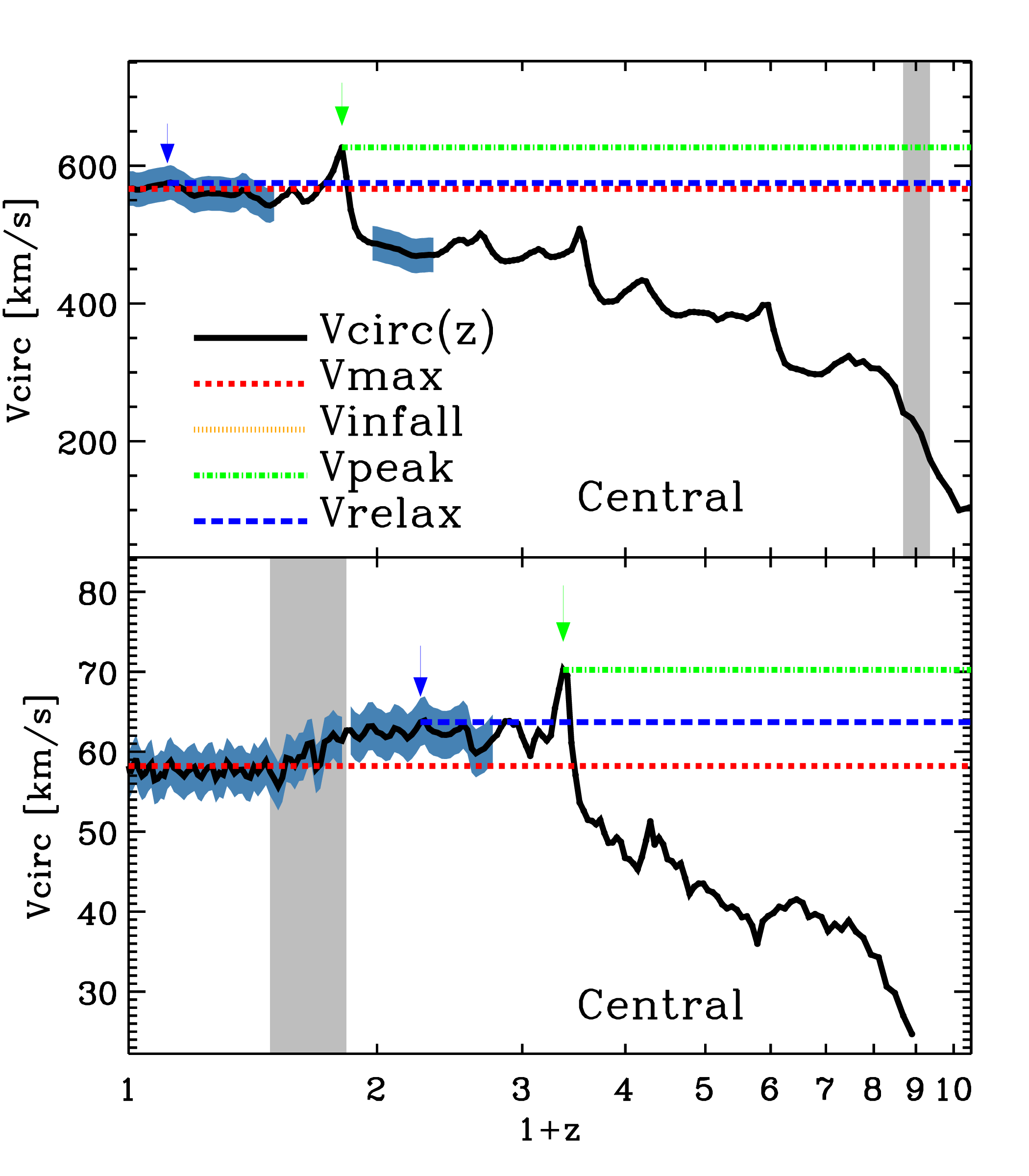}
\includegraphics[width=0.475\textwidth]{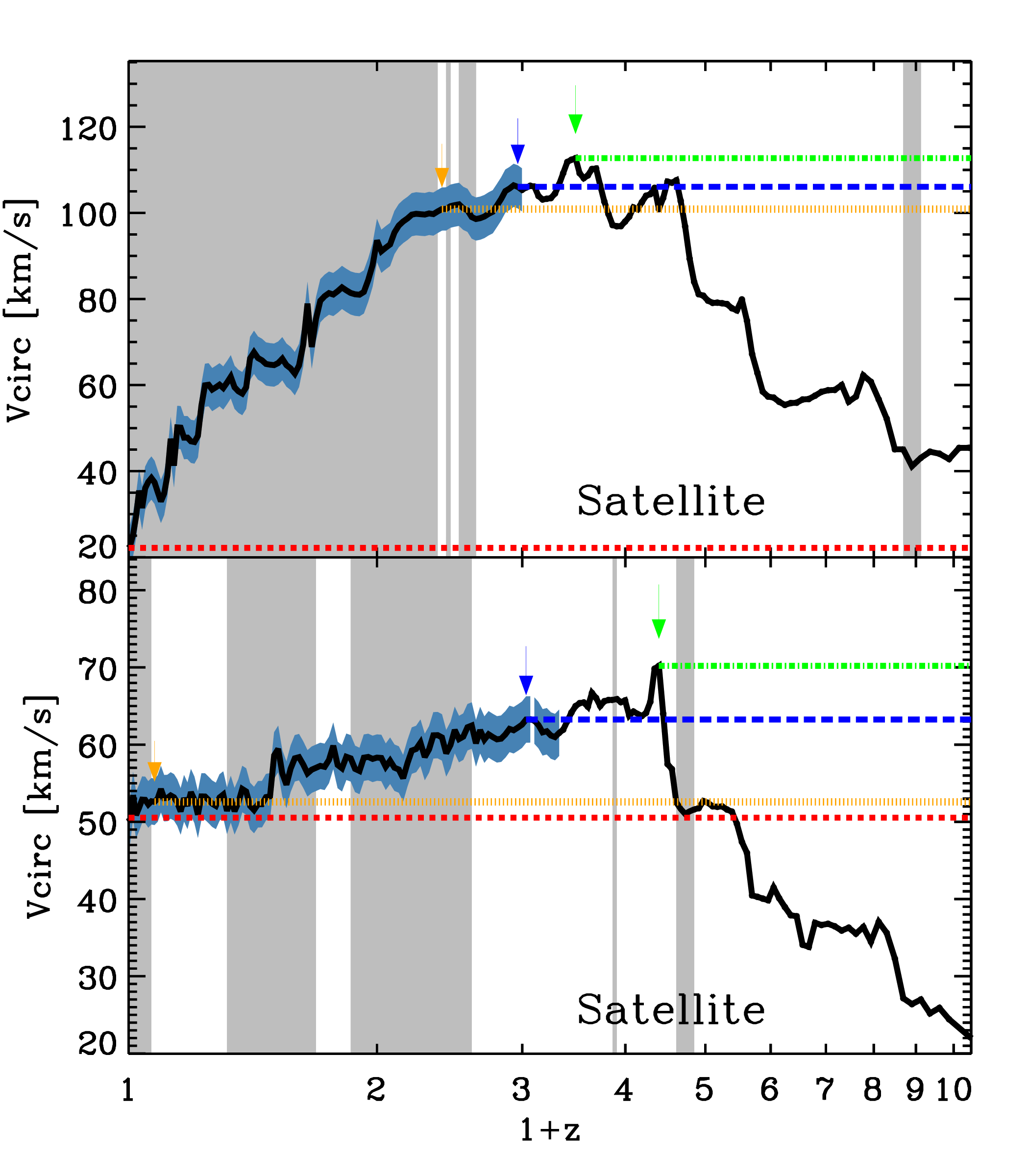}
\end{center}
\caption{
\label{fig:DefParamsC} Evolution of the maximum circular velocity of two
central (left panel) and two satellite (right panel) subhaloes in DMO.  The
black solid lines show the circular velocity, the grey coloured areas the
periods during which the subhaloes are satellites, and the blue coloured
regions the intervals during which the subhaloes satisfy our relaxation
criterion. Horizontal lines highlight the circular velocity at $z=0$ ($\Vmax$,
red dashed line), the circular velocity at the last infall for satellites and
$\Vmax$ for centrals ($\Vinfall$, orange dotted line), the maximum circular
velocity that a subhalo has had ($\Vpeak$, green dot-dashed line), and the
maximum circular velocity that a subhalo has reached while it satisfied our
relaxation criterion ($\Vrelax$, blue long dashed line).}
\end{figure*}

\subsection{SHAM flavours}

The main assumption of SHAM is that there is a one-to-one relation between a
property of a DM subhalo and a property of the galaxy that it
hosts. The galaxy property is usually taken to be the stellar mass (or K-band
luminosity), since this is expected to be tightly correlated with the DM
content of the host halo (contrary to e.g. the star formation rate, which could
be more stochastic). The subhalo property should capture the time-integrated
mass of gas available to fuel star formation, but there is no consensus as to
what the most adequate subhalo property is\footnote{Properties used in the
literature include $M_{\rm DM}$ \citep{2004MNRAS.353..189V,2006ApJ...643...14S},
maximum circular velocity at present for centrals and at infall for satellites
\citep{2006ApJ...647..201C}, virial mass for centrals and mass at infall for
satellites \citep{2010MNRAS.403.1072W,2010ApJ...717..379B}, virial mass for
centrals and the highest mass along the merger history for satellites
\citep{2010ApJ...710..903M}, and highest circular velocity along the merger
history \citep{2011ApJ...742...16T,2013MNRAS.432..743N} \citep[see ][for a detailed comparison between the previous properties]{2013ApJ...771...30R}.}.

A commonly used property in SHAM is the maximum of the radial circular velocity 
profile (which
can be regarded as a measure of the depth of the potential well of a subhalo)
defined at a suitable time:

\begin{equation}
\Vcirc(z) \equiv \text{max} \Big[\sqrt{GM(z,<r)/r}\Big].
\end{equation}

\noindent where  $M(<r)$ is the mass enclosed inside a radius $r$. 

There are several reasons to prefer circular velocity over halo mass in SHAM: i)
it is typically reached at one tenth of the halo radius, so it is a better
characterization of the scales that we expect to affect the galaxy most
directly; ii) it is less sensitive to the mass stripping that a halo/subhalo
experiences after it has been accreted by a larger object
\citep{2003ApJ...584..541H,2004ApJ...609..482K,2005ApJ...618..557N,2008ApJ...672..904P};
iii) it does not depend on the definition of halo/subhalo mass.

However, the $\Vcirc(z)$ of DM objects are complicated functions, which can
display non-monotonic behaviour in time, with transient peaks and dips, and
that are subject to environmental and numerical effects. This is illustrated by
Fig. \ref{fig:DefParamsC}, which shows examples of the evolution of the
circular velocity for two central (left panel) and two satellite (right panel)
subhaloes in DMO. These subhaloes are selected to illustrate the evolution 
of the maximum circular velocity in typical centrals and satellites. 
We can see that there is no obvious time at which $\Vcirc(z)$ should be computed 
for an accurate SHAM.

We will implement four ``flavours'' of SHAM, each using $\Vcirc(z)$ defined at a
different time: $\Vmax$, $\Vpeak$, $\Vinfall$, and $\Vrelax$ (each marked by
horizontal lines and arrows of a different colour in Fig. \ref{fig:DefParamsC}).
The first three flavours have been used previously in the literature, whereas
the fourth is first used in this work. We discuss the four SHAM flavours next.

\begin{itemize}

\item[1)] $\Vmax$ is the maximum circular velocity of a subhalo at the present
time, $V_{\rm circ}(z=0)$.

\item[2)] $\Vinfall$ is the maximum circular velocity at the last time a
subhalo was identified as a central.

\item[3)] $\Vpeak$ is the maximum circular velocity that a subhalo has reached.

\item[4)] $\Vrelax$ is the maximum circular velocity that a subhalo has reached
during the periods in which it satisfied a relaxation criterion. The criterion
we use is $\Delta t_{\rm form} > t_{\rm cross}$, following a similar approach to
\cite{2012MNRAS.427.1322L}. The motivation is that after a major merger, DM
haloes typically need of the order of one crossing time ($t_{\rm cross}=2\,
r_{200}/V_{200}= 0.2/H(z)$) to return to equilibrium. Thus, we define $\Delta
t_{\rm form}$ as the look-back time from a given redshift $z_{\rm i}$ to the
redshift where the main progenitor of a subhalo reached $3/4$ of the subhalo
mass at $z_{\rm i}$ (we tested other definitions for the formation time, from
$4/5$ to $1/2$, finding roughly the same results). The periods during which this
condition is satisfied are shown as blue shaded regions in Fig.
\ref{fig:DefParamsC}. We can compute $\Vrelax$ for more than the $99\%$ of the
subhaloes in DMO and we remove the subhaloes where $\Vrelax$ cannot be
calculated. We cannot compute $\Vrelax$ for the full sample because this
quantity is not defined for subhaloes younger than one crossing time.

\end{itemize}

Although $\Vcirc$ should generally not be affected by the stripping of the
outer layers of a halo, in the right panel of Fig.  \ref{fig:DefParamsC} we can
see that it does still evolve for satellites. The decrease in $\Vcirc(z)$ after
infall is in large part due to tidal heating, a process which reduces the
density in the inner regions of the satellites \citep{2003ApJ...582..141G,
2003ApJ...584..541H, 2004ApJ...609..482K}. The tidal heating is related to the
position of a subhalo inside its host halo, being maximum at pericentric
passages. We can see an extreme case of tidal interactions in the top right
panel, where this subhalo has lost more than 99\% of its mass since it 
became a satellite. After the last infall at $1+z\sim2.3$ (grey shaded region),
the value of $\Vcirc$ decreased by about $80\%$ in a series of steps ($z\sim
1,\,0.5,\,0.3,\,0.1,\,0.05,\,\text{and}\;0$), which indeed coincide with
pericentric passages. This implies that satellite galaxies have lower values of
$\Vmax$ than central galaxies of the same stellar mass. Thus, a $\Vmax$-based
SHAM will underestimate the fraction of satellites.

Tidal heating and stripping affect not only satellites but also ``backsplash
satellites'', i.e. centrals at $z=0$ which were satellites in the past,
reducing their circular velocity while they were inside a larger halo. An
example of this process is shown in the bottom left panel of Fig.
\ref{fig:DefParamsC}, where the circular velocity of this subhalo was reduced
by about 7\% in the period during which it was a satellite (while the mass was
reduced by 50\%). 

$\Vinfall$ is less affected by these problems. Unfortunately, this parameter
also underestimates $\Vcirc$ for satellites because tidal heating starts to act
even before a satellite is accreted by its future host halo
\citep{2004ApJ...609..482K,2013MNRAS.432..336W,2014MNRAS.439.2687W}. This can be
seen in the top (bottom) right panel Fig. \ref{fig:DefParamsC}, where the value
of $\Vcirc$ starts to decrease at $1+z\sim3.4$ ($1+z\sim4.4$) while the subhalo
is accreted at $1+z\sim2.4$ ($1+z\sim1.2$).

Additionally, there are new problems associated with $\Vinfall$. The first 
concerns satellite-satellite mergers
\citep[][]{2009MNRAS.399..983A,2009MNRAS.395.1376W}, which should increase the
mass of stars in a satellite but this is not captured by $\Vinfall$. The
second is related to the definition of $\Vinfall$; it is not clear whether
we should consider $\Vinfall$ as the circular velocity at the last infall or at
previous accretion events. We can see in the bottom right panel of Fig.
\ref{fig:DefParamsC} a satellite which has undergone several alternating
central/satellite periods, decreasing in total its circular velocity by 20\%
and its mass by 70\%. 

An alternative solution is provided by $\Vpeak$ since it can capture all
episodes during which the subhalo grows, and it is not affected by a reduction
of $\Vcirc$ due to environmental effects. However, this definition similarly has
its own problems. During periods of rapid mass accretion, DM haloes are usually
out of equilibrium \citep{2007MNRAS.381.1450N}. In particular, during major
mergers the concentration can be artificially high (this is a maximum
compression phase of halo formation), which temporarily increases the value of
$V_{\rm circ}$ \citep[e.g.][]{2012MNRAS.427.1322L,2014ApJ...787..156B}. This
effect is responsible for the peaks seen in all four panels of Fig.
\ref{fig:DefParamsC}. Although at any given time it is rare to find a halo in
this phase, the value of $\Vpeak$ will likely be assigned during one of these
phases, and will thus overestimate the depth of the potential well. In addition,
this effect makes the predictions of $\Vpeak$ dependent on the number and
intervals of the output times of a given simulation.

Here we propose a new measure, $\Vrelax$, designed to overcome the problems of
$\Vmax$, $\Vinfall$, and $\Vpeak$. It is marked by arrows and horizontal lines
of blue colour in Fig. \ref{fig:DefParamsC}. $\Vrelax$ is insensitive to tidal
heating, transient peaks, and consistently defined for centrals, satellites, and
backsplash satellites. We emphasise that it is desirable to eliminate the
aforementioned problems because they represent changes in $\Vcirc$ which are not
expected to correlate with the growth history of $\Mstar$, and will thus add
extra noise to SHAM.

\begin{figure}
\begin{center}
\includegraphics[width=0.45\textwidth]{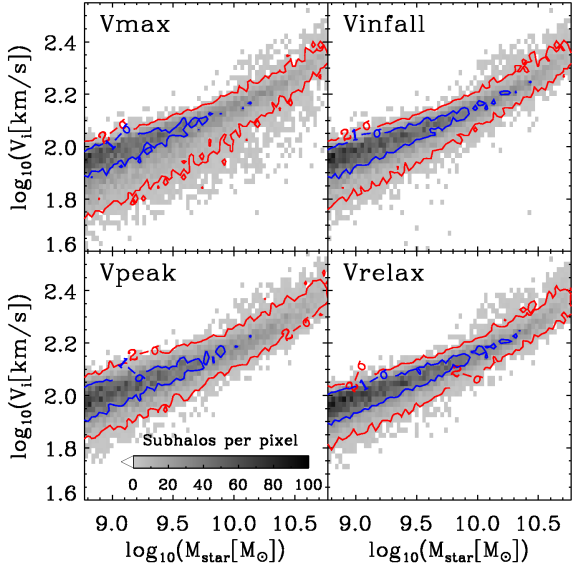}
\end{center}
\caption{
\label{fig:ScattEND} Relation between $\Mstar$ of EAGLE galaxies and SHAM
flavours for the corresponding DMO subhaloes. The grey scale represents the
number of subhaloes per pixel, which ranges from 1 (light grey) to 100 (black).
Blue and red contours mark the regions containing $68\%$ and $95\%$ of the
distribution, respectively.} \end{figure}

We now take a first look at the performance of each SHAM flavour.  Fig.
\ref{fig:ScattEND} shows the relation between each of the four properties
described above for DMO subhaloes, as indicated by the legend, and $\Mstar$ of
their galaxy counterpart in EAGLE (see \S \ref{sec:crossmatch}). All panels
show a tight correlation, which supports the main assumption of SHAM, that the
relation between stellar mass and SHAM parameters should be monotonic. However,
the scatter in the relation is different in each panel because of the effects
discussed in this section: $\Vmax$ shows the largest and $\Vrelax$ the smallest
dispersion. In the next sections we will quantify the performance of each SHAM
flavour in detail.

\subsection{Implementation}
\label{sec:scatt}

\begin{table}
\begin{center}
\caption{Parameters of the functions that fit the mean, $\mu$, and standard deviation, $\sigma$, of the model for $P(\log_{10} \Mstar[{\Msun}]|\log_{10} V_i[\kms])$. The unit of $V_i$ is $\kms$.}
\label{tabular:scatt}
\begin{tabular}{c||c|c||c|c|c|c}
\hline \hline
&\multicolumn{2}{c}{$\sigma=a+b\,\log_{10}V_i$} &\multicolumn{4}{c}{$\mu=a+b\,\tan^{-1}(c+d\,\log_{10}V_i)$} \\
& $a$ & $b$ & $a$ & $b$ &$c$ &$d$\\
\hline
$\Vmax$    & 0.60 & -0.20 & 7.03 & 5.52 & -1.84 & 1.12\\
$\Vinfall$ & 0.53 & -0.16 & 7.01 & 5.52 & -1.84 & 1.12\\
$\Vpeak$   & 0.55 & -0.16 & 7.70 & 5.42 & -1.89 & 1.05\\
$\Vrelax$  & 0.59 & -0.20 & 7.14 & 5.55 & -1.86 & 1.10\\
\hline
\end{tabular}
\end{center}
\end{table}

\begin{figure}
\begin{center}
\includegraphics[width=0.45\textwidth, height=0.55\textwidth]{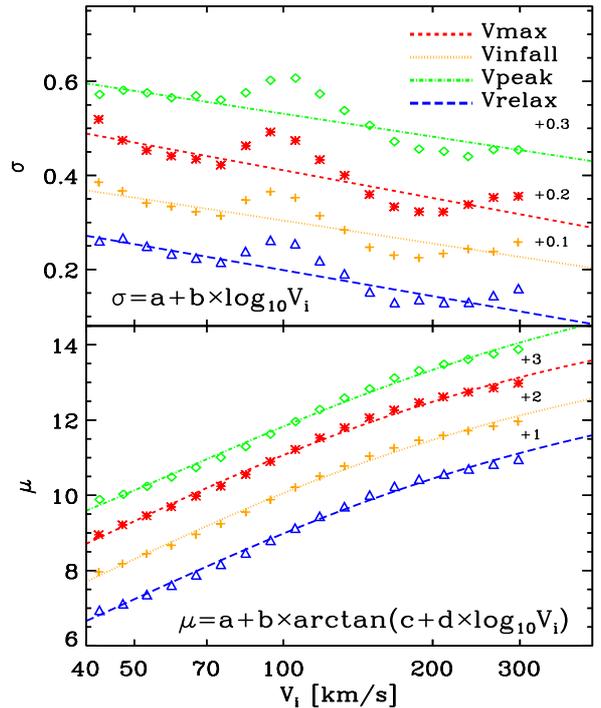}
\end{center}
\caption{
\label{fig:gausaprox} Standard deviation (top panel) and mean (bottom panel) of
the Gaussians used to fit PDFs for $\log_{10}\Mstar[\Msun]$. For clarity, we
have shifted the $\sigma$ ($\mu$) of $\Vmax$, $\Vinfall$, and $\Vpeak$ by +0.3,
+0.2, and +0.1 (+3, +2, and +1), respectively. The best fitting functions are
shown by coloured lines, and the values of the respective parameters are given
in Table \ref{tabular:scatt}.}
\end{figure}

The first step to implement the four flavours of SHAM is to compute $\probmv$:
the probability that a subhalo hosts a galaxy of mass $\Mstar$ given a certain
value of the SHAM flavour $V_i$. We compute this quantity as follows:

\begin{itemize} 

\item[1)] We select subhalo-galaxy pairs from the matched catalogues (see
\S\ref{sec:crossmatch}) with $\log_{10} \Mstar [\text{M}_{\odot}]>7$ and divide
them according to $\log_{10}V_i$ in bins of 0.05 dex. We discard bins with fewer
than 100 objects. 

\item[2)] For each $\log_{10}V_i$ bin, we compute the distribution of
$\log_{10}\Mstar$ and fit it by a Gaussian function, $G\sim
\exp(-0.5(\log_{10}{\Mstar}-\mu)^2/(\sigma)^2)$, where $\mu$ is the mean and
$\sigma$ the dispersion.

\item[3)] We fit a linear function, $\sigma=a+b\,\log_{10}V_i$, to
$\sigma(\log_{10}V_i)$ and an arctangent,
$\mu=a+b\,\tan^{-1}(c+d\,\log_{10}V_i)$, to $\mu(\log_{10}V_i)$. The values of
the best-fit parameters are given in Table \ref{tabular:scatt} and the quality
of the fit can be judged from Fig. \ref{fig:gausaprox}.

\item[4)] Using these functions, we model $\probmv$ as
$G[\mu(\log_{10}V_i),\sigma(\log_{10}V_i)]$.

\end{itemize}

Our second step is to assign a value of $\Mstar$ to every subhalo in DMO (not
only those with an EAGLE counterpart) by randomly sampling $\probmv$. This
creates a catalogue that captures the appropriate stochastic relation between
$\Mstar$ and the parameter $V_i$. If the relation for EAGLE galaxies were also
stochastic with respect to the underlying density field, then we would expect
these catalogues to have the same clustering properties as EAGLE.

\begin{figure*}
\begin{center}
\includegraphics[width=0.8\textwidth,height=0.5\textwidth]{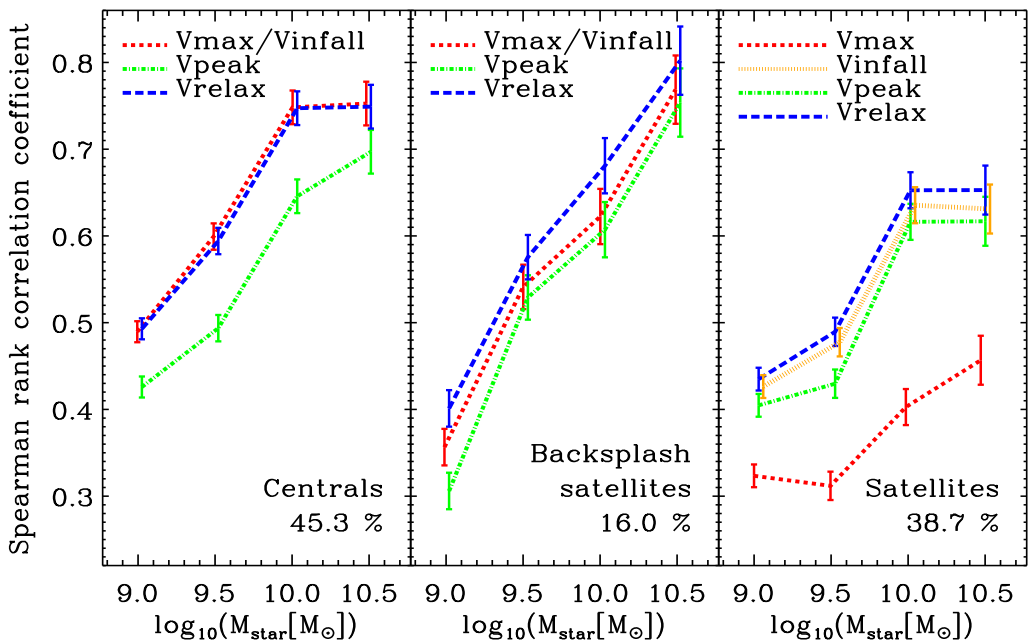}
\end{center}
\caption{
\label{fig:CorrEND} The Spearman rank correlation coefficient between the
$\Mstar$ of EAGLE galaxies and each of four parameters used to perform SHAM. 
The subhaloes are divided into three categories: centrals (left panel),
backsplash satellites (central panel), and satellites (right panel), see the
main text for more details. The fraction of objects in each category is given
in the legend. The red (orange) points are displaced horizontally by -0.03
(+0.03) dex for clarity.}
\end{figure*}

We note we have verified that the resulting stellar mass function agrees closely
with that of the EAGLE simulation. However, to ensure {\it identical} mass
functions and thus to make subsequent comparisons more direct, we assign to each
SHAM galaxy the value of $\Mstar$ of the EAGLE galaxy at the same rank order
position. Hereafter, we will refer generically to the galaxy catalogues created
in this way as ``SHAM galaxies'' and specifically to the galaxy catalogues
generated by a particular SHAM parameter as ``$V_i$ galaxies''.

We compute 100 realizations of SHAM for every flavour using different random
seeds. The results presented in the following sections are the mean of all the
realizations and the errors the standard deviation.

\section{Results}
\label{sec:results}

In this section we test how well SHAM reproduces different properties of
EAGLE galaxies. In particular, we will explore the predicted stellar mass of
individual subhaloes (\S\ref{sec:pearson}), the halo occupation distribution
(\S\ref{sec:hod}), the number density profiles inside haloes
(\S\ref{sec:profile}), the clustering in real and redshift space
(\S\ref{sec:rspace}, \S\ref{sec:zspace}), and the assembly bias
(\S\ref{sec:assembly}).

We present results for 4 bins in stellar mass, as indicated in Table
\ref{table:cross}. This range was chosen to include only well sampled and
well resolved galaxies (comprised of more than 230 star particles) and 
bins with enough
galaxies to allow statistically significant analyses (more than 1600 galaxies).

\subsection{Correlation between $\Mstar$ and $V_i$}
\label{sec:pearson}

In \S\ref{sec:SHAM} we discussed that in some cases $\Vmax$, $\Vinfall$, and
$\Vpeak$ are unintentionally affected by physical and numerical effects, which
degrades the performance of SHAM. We also argued that $\Vrelax$ does not
present any obvious problem and thus we expected it to be the SHAM flavour that
correlates most strongly with $\Mstar$. This was qualitatively supported by
Fig. \ref{fig:ScattEND}. We start this section by quantifying these statements
using the Spearman rank correlation coefficient between the $\Mstar$ of EAGLE
galaxies and the SHAM flavours of DMO subhaloes.

The Spearman coefficient measures the statistical dependence between two
quantities and is defined as the Pearson correlation coefficient between the
ranks of sorted variables. A value of unity implies a perfect
correlation, which in our case means that the stellar mass of a galaxy is
completely determined by its SHAM parameter, i.e. that the relation is
monotonic and thus without scatter. A value close to zero means that the
relation between the SHAM parameter and $\Mstar$ is essentially random.

In Fig. \ref{fig:CorrEND} we show the Spearman coefficient for the correlation
between $\Mstar$ and each of our four SHAM parameters. We divide our sample into
three groups: i) present-day central subhaloes that have been centrals for their
entire merger history except for at most 4 snapshots (centrals, left panel), ii)
present-day central subhaloes that have been satellites more than 4 snapshots in
the past (backsplash satellites, central panel), and iii) present-day satellites
(satellites, right panel). 

In general, we find that the correlation increases with $\Mstar$, that it is
stronger for centrals than for satellites, and that $\Vrelax$ displays the
strongest correlation with $\Mstar$. Regarding the different SHAM flavours, we
find that i) for centrals $\Vpeak$ produces the weakest correlation, ii) for
satellites $\Vmax$ shows the weakest correlations, and iii) $\Vinfall$ and
$\Vrelax$ consistently display the best performance, with $\Vrelax$ showing a
slight improvement over $\Vinfall$ for satellites. 

\begin{figure*}
\begin{center}
\includegraphics[width=0.75\textwidth]{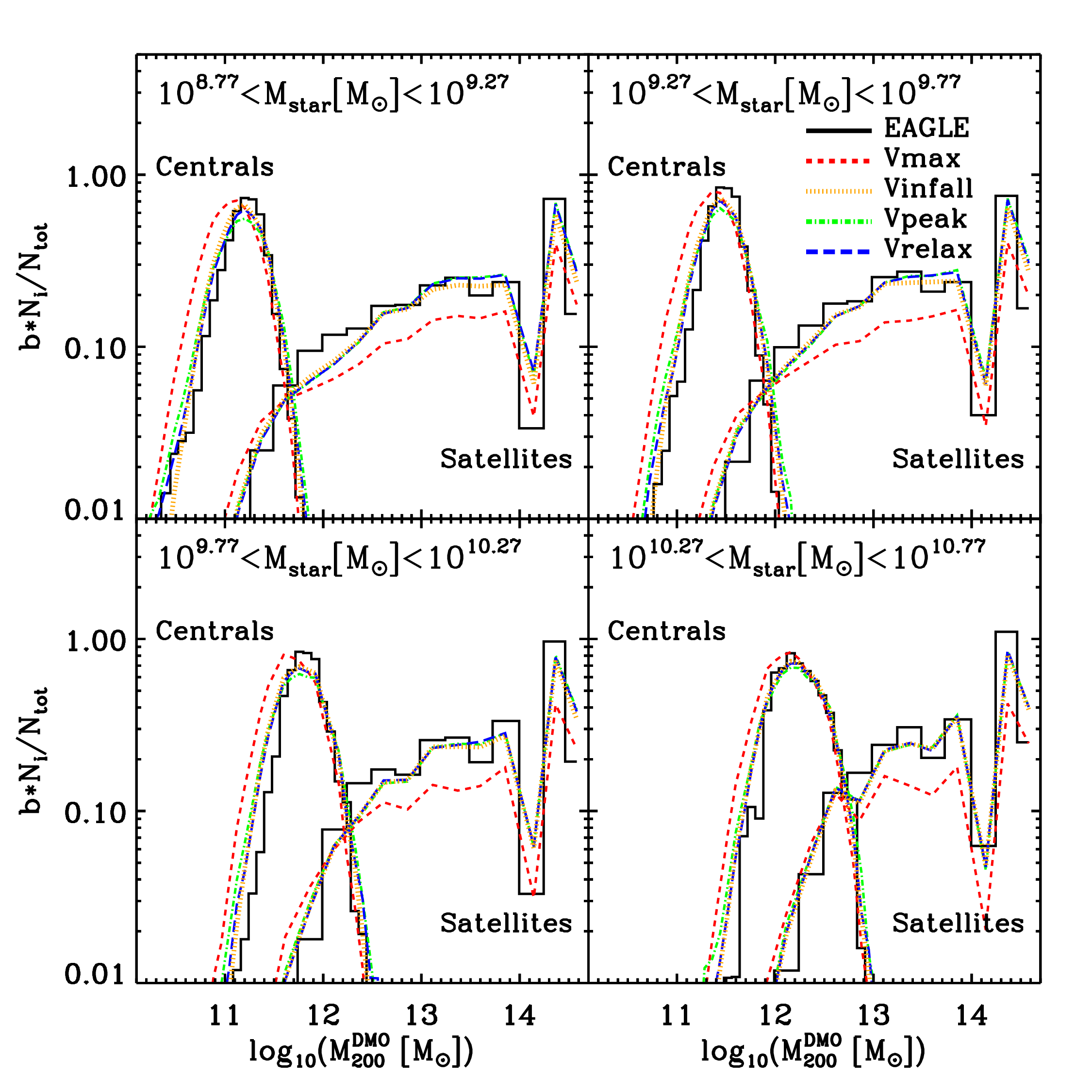}
\end{center}
\caption{
\label{fig:hod} 
The distribution of host halo masses, $M_{200}$, for SHAM and EAGLE galaxies in
different $\Mstar$ bins. Histograms show the results for EAGLE galaxies and
coloured lines for different SHAM flavours, as detailed in \S3.2. The left
(right) curves display the number ${\rm N}_{\rm i}$ of centrals (satellites) in
haloes of a given mass multiplied by the linear bias $b$ and normalized by the
total number of subhaloes ${\rm N}_{\rm tot}$. Therefore, the y-axis reflects
the relative contribution of galaxies in different host halo mass bins to the
large-scale correlation function. Note that for EAGLE galaxies we employ the
$M_{200}$ of the DMO counterpart, which makes our comparison less dependent on
the baryonic processes which might alter the mass of the host halo.}
\end{figure*}

Our results can be understood from the discussion in \S2. For centrals, $\Vmax$
and $\Vinfall$ are identical by construction and they are close to the value of
$\Vrelax$ because $\Vcirc$ tends to increase with decreasing redshift for
centrals. On the other hand, $\Vpeak$ is usually established while $\Vcirc$ is
temporarily enhanced as a result of merger events. For backsplash satellites,
$\Vmax$ and $\Vinfall$ are also identical by construction, but, unlike
$\Vrelax$, they are insensitive to their more complicated history, which
explains their weaker correlation with $\Mstar$.

Finally, satellites display the weakest correlations, with $\Vmax$ presenting
the lowest correlation coefficient. This is because $\Vcirc$ decreases soon
after infall, whereas the stellar mass can still grow until the gas is
completely exhausted (although tidal forces may strip stars).  $\Vinfall$
alleviates this problem but the interaction between the satellites and their
host haloes starts before the satellites reach the virial radii of their host
haloes \citep{2003ApJ...584..541H,2013MNRAS.430.3017B}. Because of this, 
$\Vrelax$ better captures
the expected evolution in $\Mstar$. Lastly, $\Vpeak$ is still affected by the
out-of-equilibrium artefacts discussed above. 

In sections \S4.2 and \S4.3 we will investigate how the different correlations
impact the predictions for the clustering of EAGLE galaxies. 

\subsection{The properties of SHAM galaxies}

To predict the correct galaxy clustering, SHAM has to associate galaxies with
the correct subhaloes, to allocate the right proportion of centrals and
satellites, and to place galaxies following the correct radial distribution.
Therefore, before presenting our results regarding the clustering, we will
explore these ingredients separately.

\subsubsection{Halo occupation distribution}
\label{sec:hod}

The panels of Fig. \ref{fig:hod} show the distribution of host halo masses for
centrals and satellites in different $\Mstar$ bins. The left (right) curves
display the number of centrals (satellites) in haloes of a given mass multiplied
by the linear bias\footnote{We calculate the linear bias as
$b=1+\frac{\nu^2-1}{\delta_{\rm c}}$ \citep[][]{1996MNRAS.282..347M}, where
$\delta_{\rm c}\approx1.69$ is the critical linear overdensity at collapse and
$\nu=\delta_{\rm c}/\sigma(M,z)$ is the dimensionless amplitude of fluctuations
which produces haloes of mass $M$ at redshift $z$.} expected for haloes of that
mass and normalized by the total number of subhaloes. The quantity plotted can
be interpreted as the relative contribution to the large-scale clustering from
galaxies hosted by haloes of different mass. In each panel, the histogram
presents the results for EAGLE galaxies and the coloured lines the results of
the SHAM implementations detailed in \S3.2. For EAGLE galaxies we employ the
$M_{200}$ of their host halo DMO counterpart, which makes this plot less
sensitive to baryonic effects that might systematically change the mass of DM
haloes. For the $5.1\%$ of EAGLE galaxies hosted by a halo without DMO
counterpart, we multiply $M_{200}$ by $f_{\rm DM}=1-(\Omega_{\rm b}/\Omega_{\rm
m})=0.843$. This is the average difference in $M_{200}$ between the hydrodynamic
and gravity-only EAGLE simulations, as reported by \cite{Schaller2015}.

Firstly, we see that using $\Vmax$ as SHAM parameter results in shifted
$M_{200}$ distributions and an underprediction, of about $30\;\%$, of the number
of satellites for all $\Mstar$ bins. This is a consequence of the reduction of
$\Vmax$ for satellites after being accreted, which introduces centrals hosted by
lower-mass haloes into the SHAM sample.

The distribution of EAGLE galaxies is closely reproduced by the other SHAM
implementations, for all stellar mass bins. The distributions for central
galaxies have almost identical shapes and peak at roughly the same host halo
mass. Note, however, that compared to $\Vinfall$ and $\Vrelax$, $\Vpeak$ yields
systematically broader distributions for centrals. This is consistent with the
differences in the correlation coefficient shown in the left panel of Fig.
\ref{fig:CorrEND}.

Additionally, the $\Vinfall$, $\Vpeak$, and $\Vrelax$ satellite fractions agree
to within $\sim5\%$ with those in EAGLE, although they are systematically lower,
as shown in Table \ref{tabular:hodfrac}. However, for the two lowest stellar
mass bins, there is a slight overestimate of the number of satellites in haloes
of mass $M_{200}>10^{13}\Msun$, and a somewhat larger underestimate for haloes
of mass $M_{200}<10^{13}\Msun$, as Table \ref{tabular:fracs} shows. Since the
difference is greater for the high-mass haloes, the overall satellite fraction is
underestimated. We will analyse the repercussion of these small differences in
forthcoming sections.

\begin{table}
\begin{center}
\caption{Satellite fraction for EAGLE and SHAM galaxies using $\Vmax$,
$\Vinfall$, $\Vpeak$, and $\Vrelax$.}
\label{tabular:hodfrac}
\begin{tabular}{c|c|c|c|c|c}
\hline \hline
$\log_{10}(\Mstar[\Msun])$& $\Vmax$ & $\Vinfall$ & $\Vpeak$ & $\Vrelax$ & EAGLE\\
\vspace{-8pt}\\
\cline{2-6}\\
\vspace{-16pt}\\
&\multicolumn{5}{c}{Satellite fraction}\\ \hline 
$8.77 - 9.27   $& 0.32 & 0.43 & 0.46 & 0.45 & 0.47\\
$9.27 - 9.77   $& 0.30 & 0.42 & 0.44 & 0.43 & 0.45\\
$9.77 - 10.27  $& 0.28 & 0.40 & 0.41 & 0.41 & 0.44\\
$10.27 - 10.77 $& 0.25 & 0.37 & 0.38 & 0.37 & 0.40\\ 
\hline
\end{tabular}
\end{center}
\end{table}

\begin{table}
\begin{center}
\caption{Number of satellites as a function of $\Mstar$ and $M_{200}$
for EAGLE and SHAM galaxies using $\Vrelax$.}
\label{tabular:fracs}
\begin{tabular}{c|c|r|r}
\hline \hline
$\log_{10}(\Mstar[\Msun])$ & $\log_{10}(M_{200}[\Msun])$& EAGLE & \Vrelax\\
\vspace{-8pt}\\
\cline{3-4}\\
\vspace{-16pt}\\
&&\multicolumn{2}{c}{N. of satellites}\\ \hline 
$8.77 - 9.27$   &$11.6-12.6$& 1060 & 780 \\
                &$12.6-13.6$& 1274 & 1328\\
                &$13.6-14.6$& 945  & 1057\\ \hline
$9.27 - 9.77$   &$11.6-12.6$& 584  & 444 \\
                &$12.6-13.6$& 834  & 838 \\
                &$13.6-14.6$& 633  & 695 \\ \hline
$9.77 - 10.27$  &$11.6-12.6$& 293  & 208 \\
                &$12.6-13.6$& 495  & 482 \\
                &$13.6-14.6$& 459  & 452 \\ \hline
$10.27 - 10.77$ &$11.6-12.6$& 65   & 61  \\
                &$12.6-13.6$& 280  & 253 \\
                &$13.6-14.6$& 307  & 292 \\ \hline
\end{tabular}
\end{center}
\end{table}

\begin{figure}
\begin{center}
\includegraphics[width=0.48\textwidth]{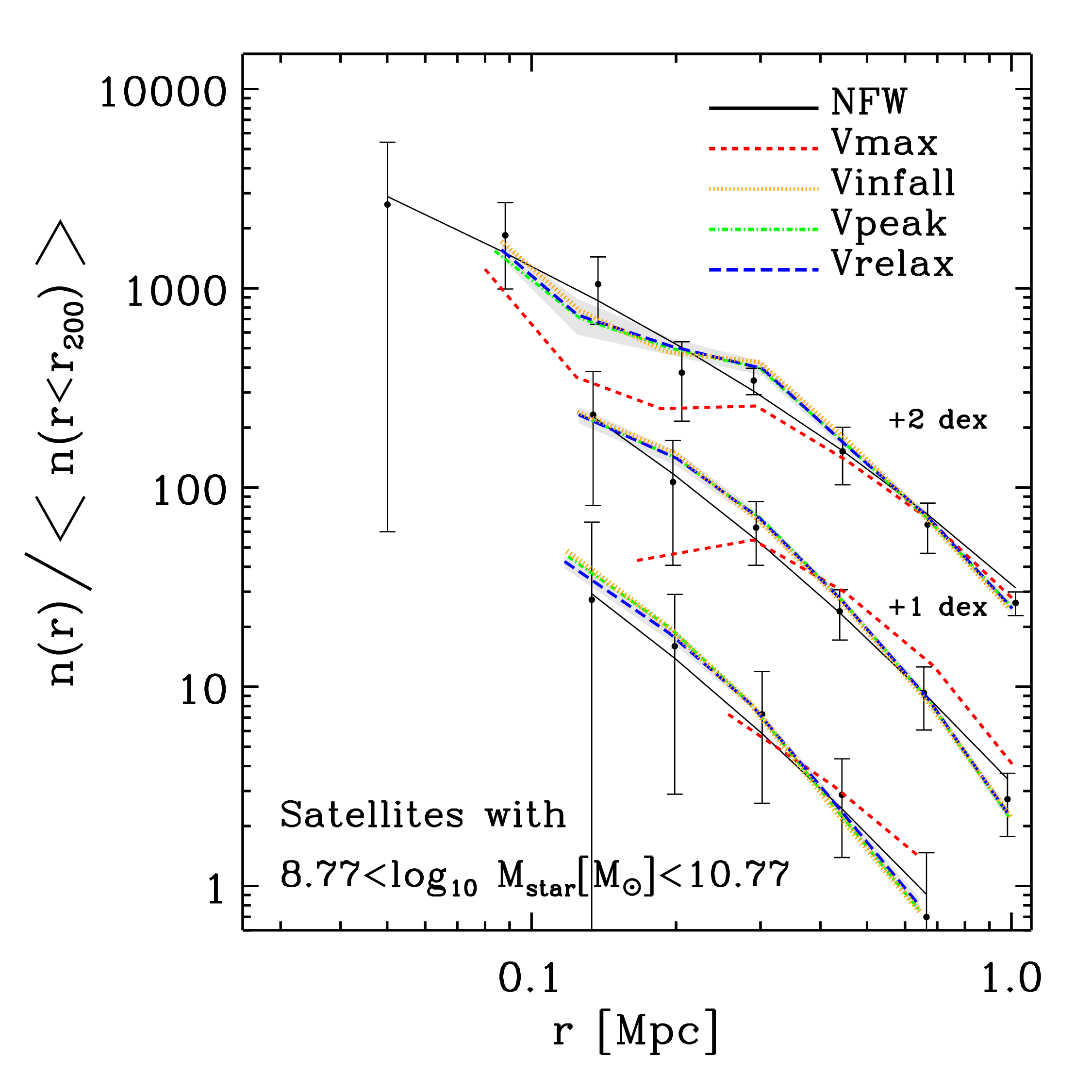}
\end{center}
\caption{
\label{fig:profile} The radial distribution of galaxies with
$8.77<\log_{10}(\Mstar [\Msun])<10.77$, inside haloes of mass
$10^{13.0}-10^{13.5} \Msun$, $10^{13.5}-10^{14.0} \Msun$ (displaced by +1 dex),
and more massive than $10^{14.0} \Msun$ (displaced by +2 dex). We present the
spherically averaged number density, normalized to the mean number density
within the host halo. Black symbols show the results for EAGLE galaxies,
whereas coloured lines show stacked results from 100 realizations of SHAM using
$\Vmax$, $\Vinfall$, $\Vpeak$, and $\Vrelax$. The error bars indicate the
$1\;\sigma$ scatter for EAGLE galaxies.  The shaded region marks the standard
deviation of 100 realizations of SHAM using $\Vrelax$. We overplot the NFW
profiles (with $r_{\rm s} = 0.81,\,0.29,\,0.21\;\Mpc$ from the most to the
least massive halo sample) that best fit the EAGLE data points shown.}
\end{figure}

\subsubsection{Radial distribution of satellites}
\label{sec:profile}

Fig. \ref{fig:profile} shows the spherically averaged number density profiles
of satellite galaxies with $8.77<\log_{10}\Mstar[\Msun]<10.77$, normalized to
the mean number density within $r_{200}$. We show results for galaxies inside
haloes in three DMO halo mass bins, as indicated by the legend. The data points
represent the profiles measured using EAGLE galaxies, whereas coloured lines
display the stacked results for SHAM galaxies. For comparison, we also plot the
best-fit NFW profile to the EAGLE data, which appears to be a good description
over the range of scales probed. 

Given the statistical uncertainties, the number density profiles of EAGLE and
SHAM galaxies agree reasonably well with the exception of $\Vmax$. For $\Vmax$,
the differences are greater, it predicts shallower profiles and a lack of
objects in the inner parts compared to EAGLE. This is consistent with the
effects described previously: the inner parts of haloes experience large tides
and are also populated by the oldest subhaloes. In contrast, on scales
$r>0.1\;\Mpc$, the $\Vpeak$, $\Vinfall$ and $\Vrelax$ profiles are consistent
with the measurements from EAGLE for all three halo mass bins. 

\begin{figure*}
\begin{center}
\includegraphics[width=0.75\textwidth]{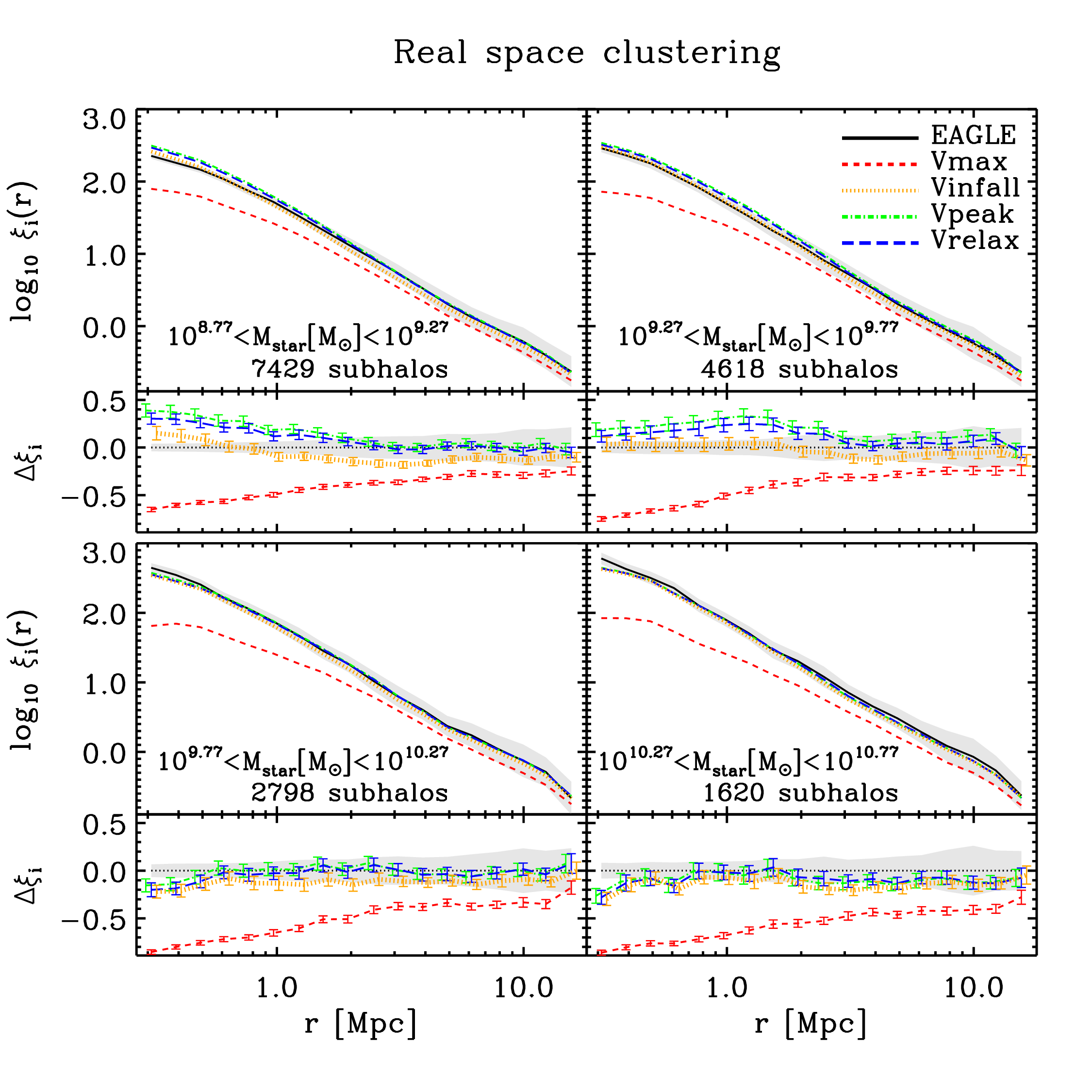}
\end{center}
\caption{
\label{fig:2pcf} Real-space two-point correlation function for galaxies in
different stellar mass bins. The black solid line shows the clustering in EAGLE,
with the grey shaded region the jackknife statistical error. The coloured lines
show the clustering predictions of SHAM using $\Vmax$ (red dashed), $\Vinfall$
(orange dotted), $\Vpeak$ (green dot-dashed), and $\Vrelax$ (blue long dashed).
The error bars indicate the standard deviation of 100 realizations of SHAM for
each flavour. In the lower half of each panel we display the relative difference
of SHAM with respect to EAGLE ($\Delta\xi_i=\xi_i/\xi_{\rm EAGLE}-1$). Note that
the green and orange lines are slightly displaced horizontally for clarity.
Using $\Vrelax$ as SHAM parameter, we retrieve the clustering of EAGLE galaxies
to within 10\% on scales greater than $2\;\Mpc$.}
\end{figure*}

\begin{figure*}
\begin{center}
\includegraphics[width=0.75\textwidth]{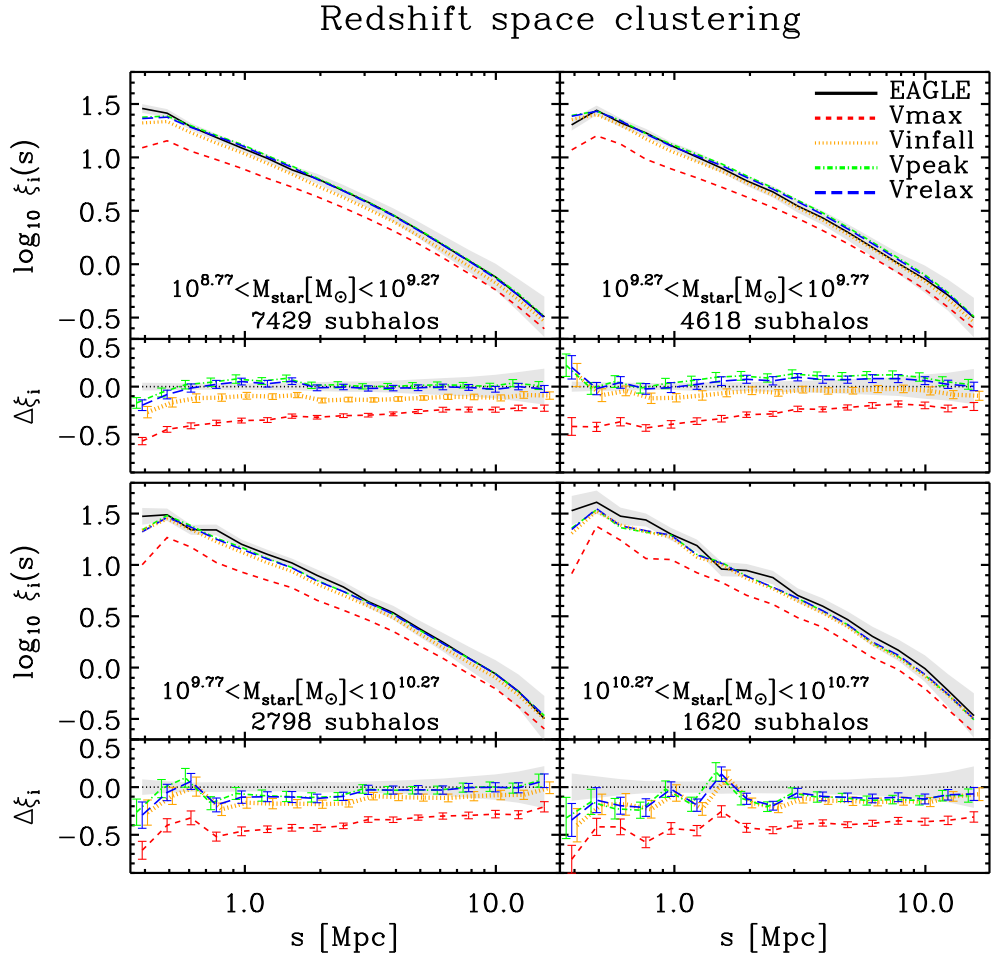}
\end{center}
\caption{
\label{fig:rsd} 
Same as Fig. \ref{fig:2pcf} but for correlation functions computed in redshift
space. The agreement between the clustering of EAGLE galaxies and $\Vpeak$ and
$\Vrelax$ galaxies is even better in redshift space than in real space for the
two lowest stellar mass bins. The main reason of the improvement on small scales
is that most of the galaxies separated by those scales in redshift space are at
larger distances in real space, where $\Vpeak$ and $\Vrelax$ galaxies accurately
reproduce the clustering of EAGLE galaxies.}
\end{figure*}

\begin{figure*}
\begin{center}
\includegraphics[width=0.75\textwidth]{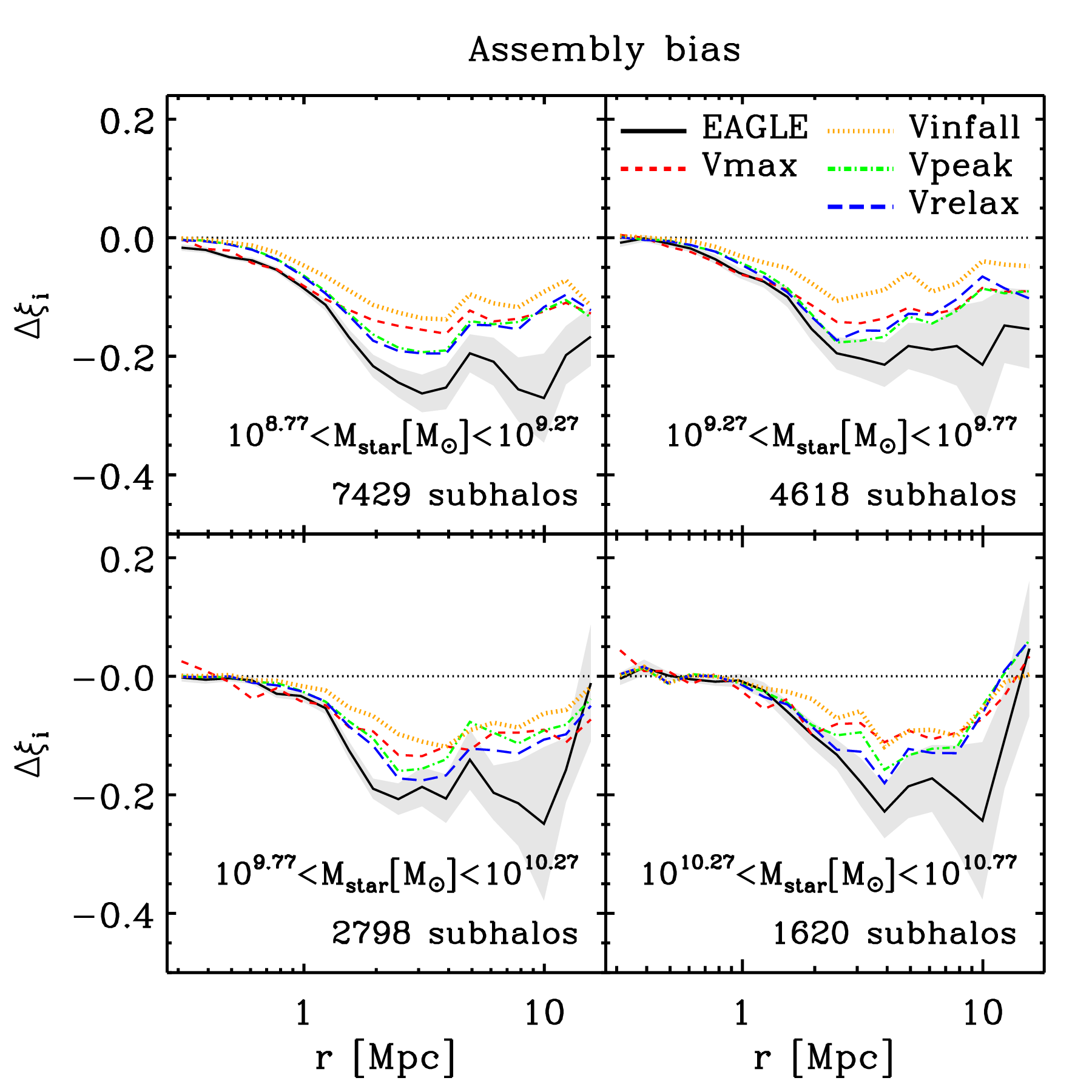}
\end{center}
\caption{
\label{fig:shuff}
The relative difference of the 2PCFs of galaxies to that of a catalogue where
galaxies are shuffled among haloes of the same mass ($\Delta \xi_i=\xi^{\rm
shuff}_i/\xi^{\rm orig}_i-1$ , see \S\ref{sec:assembly} for more details). We
adopt the same labelling as in Fig. \ref{fig:2pcf}. The grey shaded areas show
the standard deviation after applying the shuffling procedure 100 times for
EAGLE galaxies.} 
\end{figure*}

\subsection{Galaxy clustering}
\label{sec:correl}

We are now in the position to investigate the performance of SHAM in predicting
the clustering of galaxies. We first discuss the two-point correlation
function (2PCF) in real-space (\S\ref{sec:rspace}), then the monopole of the
redshift-space correlation function (\S\ref{sec:zspace}), and we end with
an exploration of assembly bias in both EAGLE and SHAM (\S\ref{sec:assembly}).

We compute the 2PCF, $\xi(r)$, by Fourier transforming the galaxy number density
field, which is a faster alternative to a direct pair count. We provide details
of the procedure in Appendix B. We estimate the statistical uncertainties in the
2PCF of EAGLE galaxies using a spatial jack-knife resampling
\citep[e.g.,][]{2005ApJ...630....1Z}. Summarizing, we divide the simulation box
in 64 smaller boxes and then we compute 64 2PCFs removing one of the small boxes
each time. The statistical errors are the standard deviation of the 64 2PCFs. On
the other hand, we assign errors to the 2PCF of SHAM galaxies by computing the
standard deviation of 100 realizations for each SHAM flavour.

\subsubsection{Real-Space Correlation Function}
\label{sec:rspace}

In Fig. \ref{fig:2pcf} we compare the 2PCF for EAGLE galaxies (black solid
line) with results of stacking 100 realizations of SHAM for
different stellar mass bins. In the bottom panel of each subplot, we display
the relative difference of the 2PCFs of each $V_i$ galaxy sample and EAGLE
($\Delta\xi_i=\xi_i/\xi_{\rm EAGLE}-1$).

Fig. \ref{fig:2pcf} shows that $\Vmax$ clearly underestimates the clustering on
small scales, which is consistent with the underestimation of the satellite
fraction discussed earlier. A lower satellite fraction also implies a lower
mean host halo mass and a smaller bias, which explains the underestimation of
the correlation function on larger scales.

On the other hand, $\Vinfall$, $\Vpeak$, and $\Vrelax$ galaxies agree very
closely with the EAGLE measurements. On scales greater than 2 $\Mpc$, all three
flavours are statistically compatible with the full hydrodynamical results. We
note that the small differences are of the same order as the variance
introduced by different samplings of $\probmv$. For the two higher stellar mass
bins, the statistical agreement is extended down to $400\;\kpc$.

For the two lower stellar mass bins, we measure statistically significant
differences on small scales, especially for $\Vpeak$ and $\Vrelax$ galaxies.
The SHAM clustering appears to be $20-30\;\%$ high, which could originate from
either more concentrated SHAM galaxy distributions inside haloes, or from an
excess of satellite galaxies. At first sight, the latter explanation appears to
contradict our previous finding that the satellite fraction is underpredicted
by SHAM. However, the small-scale clustering will be dominated by satellites
inside very massive haloes\footnote{For instance, in the case of the
small-scale clustering of galaxies in the lowest stellar mass bin, the
contribution of satellites inside haloes with $M_{200}>10^{13}\Mass$ is almost
an order of magnitude larger than that of satellites in haloes with
$M_{200}<10^{13}\Mass$.}, whose number is indeed overpredicted (c.f. Table
\ref{tabular:fracs}).

Additionally, Fig. \ref{fig:hod} showed that $\Vinfall$ resulted in the same
underestimation of the overall satellite fraction as $\Vpeak$ and $\Vrelax$ but
a somewhat smaller satellite fraction in the high halo mass range. This 
explains the weaker small-scale clustering seen in Fig. \ref{fig:2pcf} and
consequently the slightly better agreement with EAGLE. Note, however, that the
smaller number of satellites could be caused by the fact that $\Vcirc$
decreases even before accretion, especially near very massive haloes.  This
suggests that the apparent improved performance of $\Vinfall$ could be simply a
coincidence. We will investigate these hypotheses in \S5.

\subsubsection{Redshift-space Correlation Function}
\label{sec:zspace}

Fig. \ref{fig:rsd} is analogous to Fig. \ref{fig:2pcf} but for the
redshift-space 2PCFs. We compute 2PCF in redshift-space because they are more
directly comparable with observations than the 2PCF in real-space. We transform
real- to redshift-space coordinates ($\mathbf{r}$ and $\mathbf{s}$,
respectively) in the plane-parallel approximation: $\mathbf{s}=\mathbf{r}+(1+z)
(\mathbf{v} \cdot \hat{\mathbf{k}})/H(z)$, where $\mathbf{v}$ the peculiar velocity,
$H(z)$ is the Hubble parameter at redshift $z$, and $\hat{\mathbf{k}}$ is the
unit vector along the $z$ direction. On scales greater than $6\;\Mpc$, this
transformation enhances the clustering signal due to the Kaiser effect
\citep{1987MNRAS.227....1K}. On smaller scales, motions inside virialised
structures produce the so-called finger-of-god effect, smoothing the correlation
function.

The differences between the SHAM flavours are qualitatively similar in real and
redshift space: $\Vmax$ underpredicts the clustering on all scales and for all
$\Mstar$ bins, the remaining SHAM flavours are statistically compatible with
EAGLE on scales $\gtrsim 1\;\Mpc$, and the clustering amplitude of $\Vinfall$
is systematically below that of $\Vrelax$ and $\Vpeak$. On the other hand,
compared with the real-space 2PCFs, there is better agreement between
$\Vrelax$, $\Vpeak$ and EAGLE on small scales for the two lowest mass bins.
This improvement is likely a result of two effects. First, a considerable
fraction of close pairs in redshift space will be much further apart in real
space, and hence better modelled by SHAM. Second, the incorrect HOD that SHAM
galaxies show can be compensated by a stronger smoothing of the 2PCF: a greater
number of satellites in high-mass haloes would increase the small-scale
clustering, but these satellites would also have a higher velocity dispersion.

If the agreement between SHAM and EAGLE galaxies were reached because of the
cancellation of different sources of error, then this would impact other orthogonal
statistics, for instance, the strength of the so-called assembly bias (other
examples are the high-order multipoles of the redshift space 2PCF). We explore
this next.

\subsubsection{Assembly bias}
\label{sec:assembly}

Assembly bias generically refers to the dependence of halo clustering on any
halo property other than mass, such as formation time, concentration, or spin
\citep[see, e.g.,][]{2005MNRAS.363L..66G, 2007MNRAS.377L...5G}. It has been
robustly detected in DM simulations, but it is not clear what is the effect of
assembly bias on galaxy clustering. This is because a given galaxy sample will
typically be a mix of haloes of different masses and properties. Although the
strength of the effect depends on the assumptions of the underlying galaxy
formation model, semi-analytic galaxy formation models and SHAM both suggest
that assembly bias is indeed important \citep{2007MNRAS.374.1303C,
2014MNRAS.443.3044Z, 2014arXiv1404.6524H}. To our knowledge, this issue has not
yet been investigated with hydrodynamical simulations.

In this section we explore whether assembly bias is present in EAGLE and
whether the different SHAM flavours are able to predict its amplitude. To
quantify the effect, we will compare SHAM and EAGLE 2PCFs to those measured in
shuffled galaxy catalogues, which are built following the approach of
\cite{2007MNRAS.374.1303C}:

\begin{itemize}

\item[1)] We compute the distance between each satellite galaxy and the
 centre-of-potential (COP) of its host halo. This distance is by definition zero
 for central galaxies.

\item[2)] We bin haloes according to $M_{200}$ using a bin size of 0.04 dex.
We verified that our results are independent of small changes in the bin
widths.

\item[3)] We randomly shuffle the entire galaxy population between haloes in
the same mass bin. 

\item[4)] Finally, we assign a new position to each galaxy by moving the galaxy
away from the COP of its new halo by the same distance that we calculated in 1).  

\end{itemize}

Fig. \ref{fig:shuff} shows the mean relative difference between 100 realizations
of the shuffled catalogues and the original for different bins of stellar mass.
The black solid lines display the results for EAGLE galaxies and the coloured
lines for SHAM galaxies. Since the position of galaxies/subhaloes is independent
of the environment in the shuffled catalogues, their clustering should depend
exclusively on the host halo mass. Therefore, any deviations from zero in Fig.
\ref{fig:shuff} can be attributed to the assembly bias. Note that on small
scales the ratio goes to zero by definition since the shuffling procedure does
not alter the clustering of galaxies inside the same halo\footnote{Note that our
findings would remain nearly the same if instead we shuffled centrals and
satellites separately following \cite{2014MNRAS.443.3044Z}. This is because
centrals and satellites with the same $\Mstar$ rarely reside in the same halo
(see Fig. \ref{fig:hod}).}.

We can clearly see that all shuffled catalogues underestimate the clustering
amplitude for $r\gtrsim 1\;\Mpc$. In the case of EAGLE galaxies, the differences
are $\sim20\%$ on scales greater than $2\;\Mpc$, roughly independent of stellar
mass. This implies that assembly bias increases the clustering amplitude
expected from simple HOD analyses by about $1/0.8 = 25\;\%$. 

For SHAM galaxies, the effect goes in the same direction but is somewhat weaker
for all stellar masses (although it is more statistically significant for the
lowest mass bins). This can be interpreted as SHAM lacking some environmental
dependence of the relation between $\Mstar$ and $V_i$. Likely candidates are
tidal stripping of stars, and/or tidal stripping, harassment, and starvation
happening before a galaxy is accreted into a larger DM halo. These effects are
important because the efficiency with which a given halo creates stars will
depend on the large-scale environment. We will return to these issues in the
next section.
 
Before closing this section, it is interesting to note the particular case of
$\Vinfall$, which was the SHAM flavour that agreed best with the real space
2PCF of EAGLE data. The fact that the strength of the assembly bias is roughly
a factor of two smaller than in EAGLE supports the idea that the previous
agreement was partly coincidental. Since $\Vinfall$ will be reduced near large
haloes due to interactions experimented by subhaloes before being accreted, the
number of satellites will decrease and the 2PCF will decrease on small scales.
However, this will likely occur for the wrong haloes, which will result in a
misestimated amplitude for the assembly bias. 

\section{Testing the assumptions underlying SHAM}
\label{sec:discuss}

In the previous section we showed that SHAM reproduces the clustering of EAGLE
galaxies to within $10\;\%$ on scales greater than $2\;\Mpc$ and the corresponding
assembly bias reasonably well. However, small differences remain, most notably
the clustering on small scales and the strength of assembly bias. In this
section, we will directly test four key assumptions behind SHAM with the aim of
identify the likely cause of the disagreement. Unless stated otherwise, we will
employ $\Vrelax$.

\subsection{Assumption I: The relation between $\Mstar$ and $V_i$ is
independent of redshift}

\begin{figure}
\begin{center}
\includegraphics[width=0.45\textwidth]{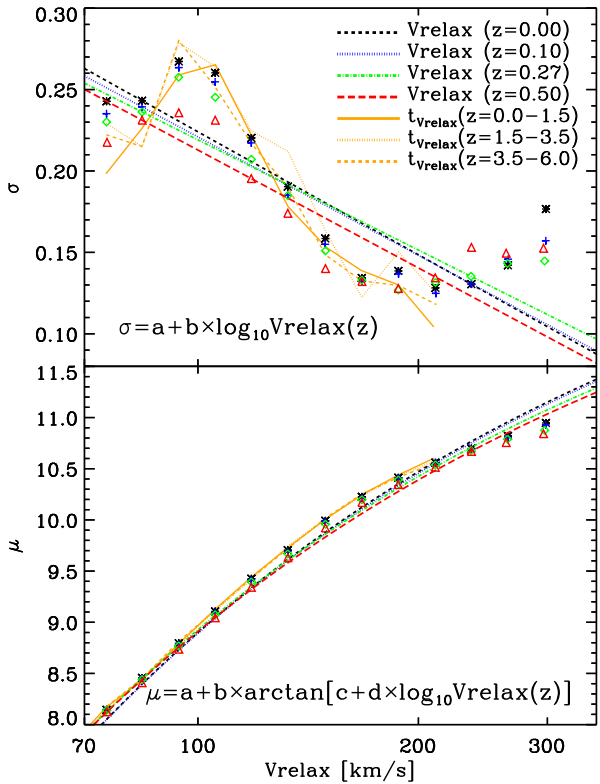}
\end{center}
\caption{
\label{fig:gausszz} Standard deviation (top panel) and mean (bottom panel) of
the Gaussian functions used to fit the dependence of the stellar mass PDF on
$\Vrelax$ at different redshifts. The symbols represent the measurements of the
widths and the centres and the lines show the fits. Neither the scatter nor the
mean of $\Mstar$ and $\Vrelax$ evolves significantly. The orange lines show the
results for galaxies at $z=0$ that have reached $\Vrelax$ at $z=0-1.5$ (solid),
$z=1.5-3.5$ (dotted), and $z=3.5-6$ (dashed).}
\end{figure}

\begin{figure*}
\begin{center}
\includegraphics[width=0.75\textwidth]{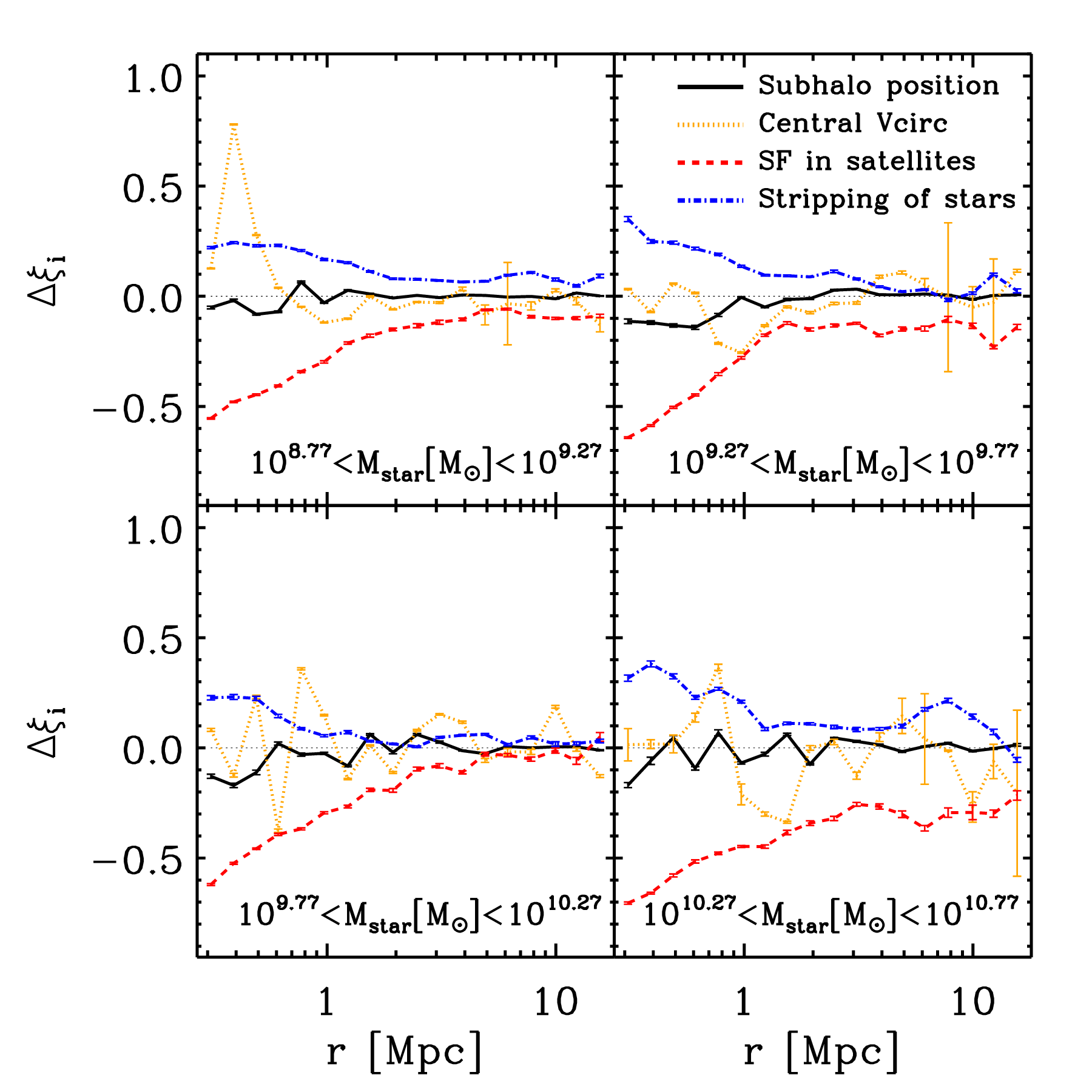}
\end{center}
\caption{
\label{fig:test} The impact on the 2PCF of different assumptions made by SHAM.
Different lines compare the 2PCF of EAGLE with those of catalogues that aim to
isolate different physical effects not included in SHAM in order to quantify
their importance for modelling galaxy clustering. Black solid lines show the
impact of baryonic effects on subhalo positions. Orange dotted lines show the
impact of baryonic effects on $\Vcirc$. Red dashed lines assess the importance
of star formation in satellites after accretion. Blue dot-dashed lines show the
impact of the stripping of stars inside massive haloes. The error bars display
the jackknife statistical errors. See the main text for more details.}
\end{figure*}

One of the main assumptions in our implementation of SHAM is that $\Mstar$
depends on the value of $\Vrelax$, but {\it not} on the redshift at which 
$\Vrelax$ was acquired. If this were not the case, we would expect an
additional dependence on, for instance, the formation time of DM haloes. Such a
redshift dependence would be particularly important for satellites, since on
average they reach their value of $\Vrelax$ at higher redshifts than centrals.   

To test this assumption, we cross-matched the DMO and EAGLE catalogues at
redshifts $z = [0, 0.1, 0.27,$ and $0.5]$. We do this by assuming that the link
between a pair of EAGLE-DMO structures matched at $z=0$ carries over to their
main progenitors at all higher $z$. Then, we construct $\probmv$ at each
redshift, which we fit by Gaussian functions with mean $\mu$ and standard
deviation $\sigma$. In Fig. \ref{fig:gausszz} we show the results. We can see
that neither the mean nor the scatter in the relation show any strong signs of
redshift dependence. Nevertheless, to estimate the impact on the clustering, we
generated a new set of $\Vrelax$ galaxies at $z=0$ employing the scatter and
mean derived at different redshifts. We find that the differences in the 2PCF
are always below $1\;\%$. 

As a further test, we split the $z=0$ catalogue into 3 bins according to the
redshift at which $\Vrelax$ was reached: $[0-1.5]$, $[1.5-3.5]$, and $[3.5-6]$.
We overplot the mean and variance of these subsamples in Fig.
\ref{fig:gausszz} as orange lines, from which we see no obvious dependence on
redshift.

Therefore, we conclude that subhaloes of a given $\Vrelax$ statistically host
galaxies of the same $\Mstar$ at $z=0$, independently of the time at which their
$\Vcirc$ reached $\Vrelax$.

\subsection{Assumption II: Baryonic physics does not affect the SHAM property
of subhaloes}

It is well known that baryons modify the properties of their DM hosts 
\citep{1996MNRAS.283L..72N,2002MNRAS.333..299G,2005MNRAS.356..107R,2015arXiv150401437O}. 
Notable examples are an increase in the central density of DM haloes due to
adiabatic contraction, or the possible reduction due to feedback or episodic
star formation events. However, SHAM assumes that the relevant property is that
of the DM host in the absence of those baryonic effects. 

We estimate the impact of this assumption by comparing the 2PCFs of
central galaxies in our cross-matched catalogue, which we then rank order and
select using either $\Vmax$ from EAGLE or $\Vmax$ from their DMO counterpart.
We focus on central galaxies since $\Vmax$ behaves well for those objects and
should be directly relevant for $\Vinfall$ satellites. In addition, the
cross-matched catalogue is highly complete, with less than $8\%$ of central
galaxies being excluded (see Table \ref{table:cross}), thus we expect our
results to be representative of the full population.

In general, we find that the values of $\Vmax$ for EAGLE galaxies are $\sim5\;\%$
lower than for DMO galaxies, with a scatter of 0.08 dex. However, since the
scatter is $27\;\%$ of that of $\Mstar$ at a fixed $\Vmax$, we expect this
difference to have only a minor effect on the clustering. This is indeed what
we find. The orange dotted line in Fig. \ref{fig:test} shows the relative
difference of the 2PCFs. The curve is compatible with zero. Note that the noise
on scales below $0.5\;\Mpc$ is caused by the small number of objects at those
separations owing to the absence of satellite galaxies in this analysis.

Therefore, we conclude that baryonic effects introduce only small perturbations
in $V_i$ rank ordered catalogues and will thus only have a minor effect on SHAM
predictions. In any case, the noisiness of the curves do not enable us to
completely rule out small changes in the galaxy clustering due to the presence
of baryons.

\subsection{Assumption III: Baryonic physics does not affect the position of
subhaloes}

Another potential consequence of the presence of baryons is the modification of
the positions of the subhaloes, caused by the slightly different dynamics
induced by the different structure of the host halo.
\cite{2014MNRAS.440.2997V} found this effect to be important on scales below 1
$\Mpc$ (but negligible on larger scales). 

We quantify this effect by comparing the 2PCF of EAGLE galaxies in two cases; i)
using their actual positions, and ii) using the position of their DMO
counterparts. We show the relative difference between these two cases as a black
solid line in Fig. \ref{fig:test}. There are no deviations from zero on large
scales and the clustering is underestimated by around $5\;\%$ on small scales.
Therefore, the assumption that the presence of baryons does not modify the
orbits of the subhaloes is justified for the range of scales explored here.

\begin{figure*}
\begin{center}
\includegraphics[width=0.75\textwidth]{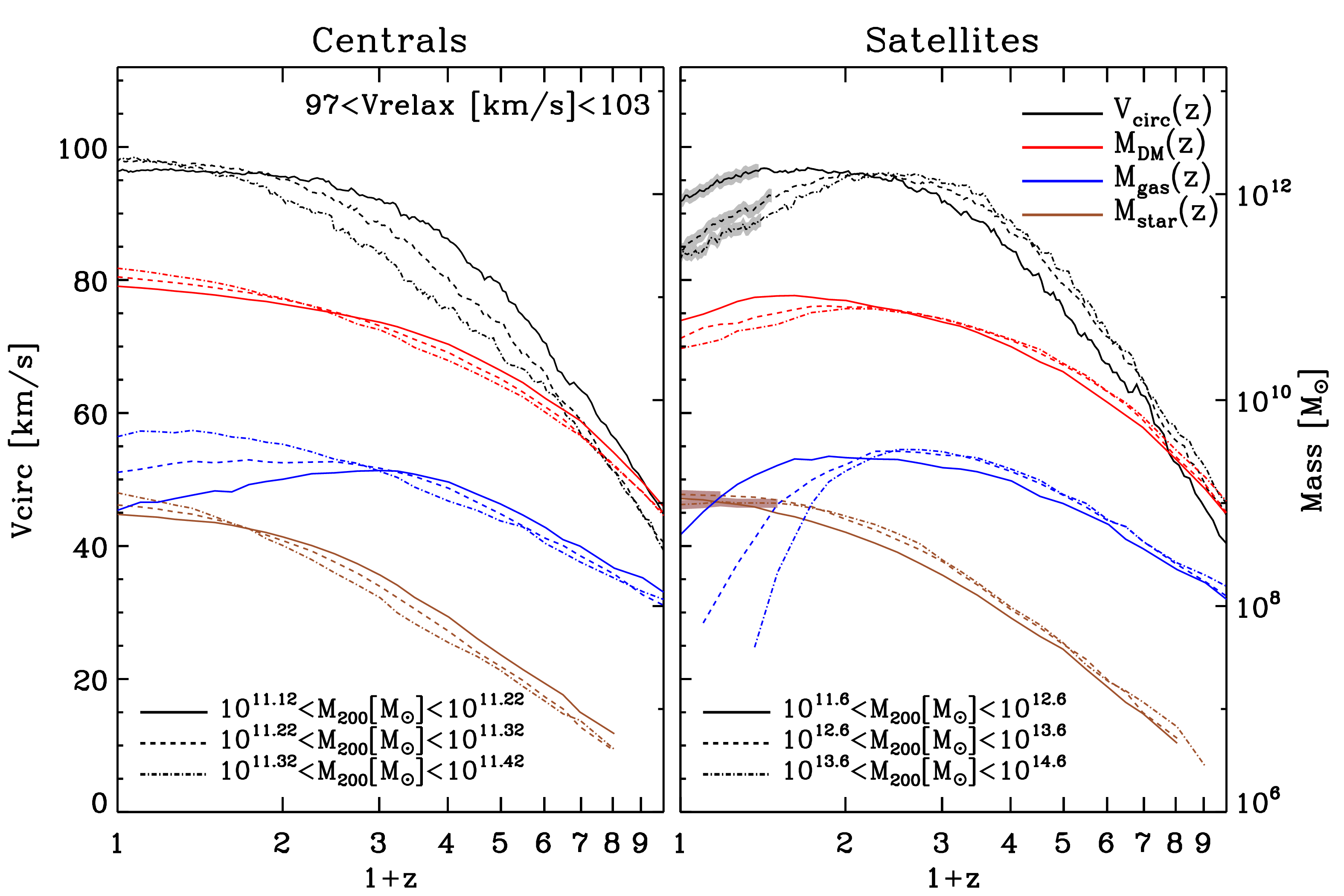}
\end{center}
\caption{
\label{fig:Vmaxevol} Evolution of the median of several subhalo properties along
the merger history for centrals (left panel) and satellites (right panel) with
$\Vrelax$ between $97$ and $103\;\kms$. The coloured lines show the evolution of
the $\Vcirc$, $M_{\rm DM}$, $\Mgas$, and $\Mstar$, as indicated by
the legend. For each component, different line styles indicate different ranges
of {\it host} halo mass. Black lines are surrounded with a grey coloured area after
$t_{\rm infall}$ and brown lines with a brown one after $t_{M_{\rm star}^{\rm
max}}$. The centrals acquire $M_{\rm star}^{\rm max}$ at $z=0$ and the ones that
reside in more massive haloes end up with higher stellar masses. For satellites
the behaviour of $\Mstar$ is more complex. After infall, the satellites which
contain gas continue forming stars until their gas is lost, but they can lose
stellar mass due to tidal stripping. The subhaloes in the right panel which
reside in haloes of
$10^{11.6}-10^{12.6},\,10^{12.6}-10^{13.6},\,10^{13.6}-10^{14.6} \,\Msun$ end up
with respectively $99,\,94,\,91\;\%$ of their $M_{\rm star}^{\rm max}$. The
stripping of DM, gas, and stars is thus more efficient for satellites in more
massive host haloes (see Table \ref{tab:stripp}).}
\end{figure*}

\subsection{Assumption IV: For a given $\Vrelax$, $\Mstar$ does not depend on
environment}
\label{sec:mtev}

We now address the assumption that deviations from the mean $\Mstar$ at fixed
$\Vrelax$ are independent of the environment. Specifically, in this subsection
we will investigate whether $\Mstar$ at fixed $\Vrelax$ is indeed uncorrelated
with the host halo mass. This is a key assumption in SHAM, because it enables
the modelling of galaxy clustering with a single subhalo property. Naturally,
the properties of galaxies are complex functions of their merger and assembly
histories, but as long as these details are not correlated with large scales,
they can be treated as stochastic fluctuations within SHAM.

We start by displaying in Fig. \ref{fig:Vmaxevol} the median growth histories of
central and satellite EAGLE galaxies within a narrow $\Vrelax$ bin from 97 to
103 $\kms$. We show the evolution of $\Vcirc$, $M_{\rm DM}$, $\Mgas$, and
$\Mstar$ for centrals (left panel) and satellites (right panel). Different line
styles indicate the results for galaxies inside three disjoint host halo mass
bins (note that the range of halo masses is different for centrals and
satellites). In the case of satellites, the grey bands mark the time after these
objects were accreted and brown bands mark the period after the maximum value
of $\Mstar(z)$ has been reached. 

Interestingly, for every parameter there is a clear distinction between
subhalos hosted by haloes of different masses. Central subhalos in the higher
host halo mass bin formed more recently, host more massive galaxies, and have
larger gas reservoirs than central subhalos hosted by less massive host haloes.
Centrals hosted by haloes in the most massive bin host a galaxie with a median
$\Mstar$ $33\;\%$ higher than the median value for all the subhalos. On the
other hand, centrals hosted by the least massive haloes have a median $\Mstar$
$18\;\%$ smaller. Therefore, the difference in $\Mstar$ is 0.22 dex and it
corresponds to $16\;\%$ of the scatter in $\Mstar$ at a fixed $\Vrelax$ (c.f.
Fig.  \ref{fig:ScattEND}), which suggests that a non-negligible fraction of the
scatter can be explained by host halo variations.

The evolution of satellites is also different in distinct host halo mass bins.
Subhalos that reside in more massive haloes reduce their $M_{\rm DM}$
and $\Vcirc$ values more significantly, suffer from stronger stripping of gas,
and stop forming stars earlier than galaxies in less massive haloes.
Furthermore, these processes appear to start prior to infall in all cases (this
also serves as an example of the limitation of $\Vinfall$), but the earlier the
higher the halo mass \citep[see
also][]{2014ApJ...787..156B,2015MNRAS.447..969B}. Nevertheless, and contrary to
the central galaxies, the final $\Mstar$ is nearly independent of the host halo
mass. It is also important to mention that the median $\Mstar$ for satellites is
$21\;\%$ higher than for centrals, which corresponds to 0.08 dex. Thus
satellite galaxies have statistically a greater $\Mstar$ than central galaxies.

In general, the evolution of satellites is more complicated than that of
centrals due to processes like strangulation, harassment, ram-pressure
stripping, and tidal stripping
\citep[e.g.][]{2010MNRAS.403.1072W,2012ApJ...754...90W}. These effects alter the
growth of satellites in a non trivial way, which is not accounted for in SHAM.
On the other hand, these processes are still not fully understood in detail, and
it is not clear how realistically current hydrodynamical simulations like EAGLE
capture them. For instance, a precise modelling of ram pressure necessarily
requires a precise modelling of the intra-cluster and interstellar medium.
Additionally, a precise modelling of tidal stripping requires precise
morphologies of the infalling galaxies. Hence, we choose to bracket their impact
on SHAM clustering predictions by considering two extreme situations.

We first consider a situation where satellite galaxies do not form or lose any
stars after infall, i.e. the value of $\Mstar$ is fixed at infall. The last
column in Table \ref{tab:stripp} compares $\Mstar$ at infall with $\Mstar$ at
$z=0$ for galaxies hosted by haloes of different masses. The corresponding
relative difference in the 2PCF is displayed by a red line in Fig.
\ref{fig:test}. In this case the satellites are less massive, which causes SHAM
to result in a $10-20\;\%$ (depending on the range of $\Mstar$ considered) lower
clustering signal on large scales. On small scales, the deficiency is larger,
reaching more than $50\;\%$. 

The second situation we consider is one where there is no tidal stripping of
stars in satellite galaxies, i.e. performing SHAM using the maximum value of
$\Mstar$ a galaxy has ever attained along its history, $\Mstar^{\rm max}$. In
Table \ref{tab:stripp} we compare the values of $\Mstar^{\rm max}$ with $\Mstar$
at $z=0$ for different bins in stellar and host halo mass. On average, we find
that the $\Mstar$ reduction begins after satellites have lost about $2/3$ of
their $M_{\rm DM}$. We also find that this effect is stronger for low mass
galaxies in higher-mass haloes, which is indeed expected due to the stronger
tides. The reduction can be up to $10\;\%$ in haloes with $M_{200} >
10^{13.6}\Mass$. On the other hand, this effect is essentially zero in haloes
with $M_{200} < 10^{12.6}\Mass$.

\begin{table}
\begin{center}
\caption{Effect of the stripping of DM and stars from satellites, and of star formation after infall. Each value corresponds to the median of the distribution and its uncertainty computed as $\sigma=1.4826\;\text{MAD}/\sqrt{n}$, where MAD is the median absolute deviation and $n$ the number of elements.}
\label{tab:stripp}
\begin{tabular}{c|c|c|c}
\hline \hline
$M_{200}[\Msun]$&$\dfrac{M_{\rm DM}}{M_{\rm DM}^{\rm max}}$&$\dfrac{\Mstar}{\Mstar^{\rm max}}$ & $\dfrac{\Mstar}{\Mstar^{\rm infall}}$\\
\hline
\multicolumn{4}{c}{$\Mstar=10^{8.77}-10^{9.27}\Msun$}\\
\hline
$10^{11.6}-10^{12.6}$& 0.428 $\pm$ 0.011 & 1.000 $\pm$ 0.000 & 1.714 $\pm$ 0.030\\
$10^{12.6}-10^{13.6}$& 0.314 $\pm$ 0.008 & 0.954 $\pm$ 0.002 & 1.828 $\pm$ 0.035\\
$10^{13.6}-10^{14.6}$& 0.274 $\pm$ 0.008 & 0.904 $\pm$ 0.004 & 1.446 $\pm$ 0.024\\                           
\hline
\multicolumn{4}{c}{$\Mstar=10^{9.27}-10^{9.77} \Msun$}\\
\hline
$10^{11.6}-10^{12.6}$& 0.458 $\pm$ 0.015 & 1.000 $\pm$ 0.000 & 1.526 $\pm$ 0.028\\
$10^{12.6}-10^{13.6}$& 0.329 $\pm$ 0.011 & 0.987 $\pm$ 0.001 & 1.752 $\pm$ 0.037\\
$10^{13.6}-10^{14.6}$& 0.278 $\pm$ 0.011 & 0.935 $\pm$ 0.004 & 1.550 $\pm$ 0.034\\
\hline
\multicolumn{4}{c}{$\Mstar=10^{9.77}-10^{10.27}\Msun$}\\
\hline
$10^{11.6}-10^{12.6}$& 0.489 $\pm$ 0.023 & 1.000 $\pm$ 0.000 & 1.360 $\pm$ 0.027\\
$10^{12.6}-10^{13.6}$& 0.352 $\pm$ 0.014 & 0.998 $\pm$ 0.000 & 1.532 $\pm$ 0.033\\
$10^{13.6}-10^{14.6}$& 0.263 $\pm$ 0.012 & 0.945 $\pm$ 0.004 & 1.433 $\pm$ 0.030\\
\hline
\multicolumn{4}{c}{$\Mstar=10^{10.27}-10^{10.77}\Msun$}\\
\hline
$10^{11.6}-10^{12.6}$& 0.670 $\pm$ 0.049 & 1.000 $\pm$ 0.000 & 1.187 $\pm$ 0.032\\
$10^{12.6}-10^{13.6}$& 0.386 $\pm$ 0.020 & 0.993 $\pm$ 0.001 & 1.197 $\pm$ 0.018\\
$10^{13.6}-10^{14.6}$& 0.238 $\pm$ 0.014 & 0.937 $\pm$ 0.005 & 1.246 $\pm$ 0.025\\
\hline
\end{tabular}
\end{center}
\end{table}

To quantify how the stripping of stars affects the SHAM clustering predictions,
we calculate the 2PCF after selecting galaxies according to $\Mstar^{\rm max}$
and compare it to our fiducial EAGLE catalogue. The result is shown by the blue
dot-dashed lines in Fig. \ref{fig:test}. In this case, the clustering is
enhanced by about $10\;\%$ on scales greater than $1\;\Mpc$ and by up to $35\;\%$ on
scales below $1\;\Mpc$. This can be understood from the fact that the satellites
are more massive, causing the satellite fraction and mean host halo mass
increase, which affects the 2PCF particularly on small scales.

The two effects considered here, stellar stripping and reduced gas supply in
satellites, affect the SHAM galaxy clustering to a similar magnitude but with
opposite sign. In particular, for all $\Mstar$ their impact is larger than the
differences between SHAM and EAGLE predictions. Thus, the final galaxy
clustering is sensitive to how these processes balance each other, which in turn
depends sensitively on baryonic processes not yet fully understood
quantitatively. On the one hand, this implies an intrinsic limitation of current
SHAM modelling that is reached when better than $\sim 20\;\%$ accuracy is
required. On the other hand, this suggests that galaxy clustering on small
scales is a powerful test for the physics implemented in hydrodynamical
simulations. For instance, if SHAM results were to be taken as the reality and
confirmed by observations, then EAGLE would implement too weak ram-pressure
stripping of massive satellite galaxies and excessive stellar stripping of
low-mass galaxies in haloes with $M_{200}> 10^{12.6} \Msun$.

\section{Conclusions}

We have used the Ref-L100N1504 EAGLE cosmological hydrodynamical simulation to
perform a detailed analysis of subhalo abundance matching for galaxies with
stellar mass ranging from $10^{8.77}\Msun$ to $10^{10.77} \Msun$. We used a
catalogue of paired EAGLE galaxies and subhaloes in a corresponding DM-only
simulation to search for an optimal implementation of SHAM, to test its
performance in terms of halo occupation numbers, radial number density profiles,
galaxy clustering, and assembly bias, and to investigate the validity of some of
the key assumptions underlying SHAM.

Our main findings can be summarized as follows:

\begin{itemize}

\item We argue that all current SHAM implementations use DM properties that are
affected by undesired physical or numerical artefacts. Thus, we propose a new
measure: $\Vrelax$, which is defined as the maximum circular velocity that a
subhalo has reached while satisfying a relaxation criterion. We also studied
SHAM using three other subhalo properties: $\Vmax$, the maximum circular
velocity at $z=0$; $\Vinfall$, the maximum circular velocity at the last time a
subhalo was a central; and $\Vpeak$, the maximum circular velocity that a
subhalo has reached. In Fig. \ref{fig:CorrEND} we show that out of the four SHAM
flavours we tested, $\Vrelax$ exhibits the strongest correlation with $\Mstar$,
independently of the subhalo history.

\item $\Vinfall$, $\Vpeak$, and $\Vrelax$ reproduce the EAGLE predictions
reasonably well (with $\Vrelax$ performing slightly better than $\Vinfall$ and
$\Vpeak$): 

\begin{itemize}

\item Fig. \ref{fig:hod} shows that the distributions of host halo masses
between EAGLE and SHAM flavours match closely. In particular, the total
satellite galaxy fraction agrees to within $5\;\%$.

\item Fig. \ref{fig:2pcf} shows that galaxy clustering strength agrees to within
$10\;\%$ on scales greater than $1\;\Mpc$ and within $30\;\%$ on smaller scales.
We highlight that this relation holds over four orders of magnitude in amplitude
and three in length scale.

\item Fig. \ref{fig:rsd} shows that in redshift space the agreement improves to
the point that there is no statistically significant discrepancy.

\item Assembly bias is present both in EAGLE and in its SHAM catalogues. Fig.
\ref{fig:shuff} shows that assembly bias increases the clustering by about
$20\;\%$.

\end{itemize}

Although small, the differences between EAGLE and SHAM are systematic and
significant. We attribute these to SHAM slightly overpredicting, compared to
EAGLE galaxies, the fraction of low-mass satellites in massive haloes.

\item Fig. \ref{fig:Vmaxevol} shows that there is a relation between $\Mstar$
and halo mass at fixed $\Vrelax$. Centrals hosted by more massive haloes
typically have higher $\Mstar$, formed more recently, and contain more gas than
those hosted by smaller haloes. Satellites that reside in more massive haloes
typically reduce their $M_{\rm DM}$ and $\Vcirc$ values more significantly,
suffer from stronger stripping of gas, and stop forming stars before accretion
and earlier than those in less massive haloes. The $\Mstar$ of satellite
galaxies at $z=0$ is independent of the host halo mass and it is $\sim 20\;\%$
greater than the $\Mstar$ of central galaxies at fixed $\Vrelax$.

\item Interactions between satellites and their host haloes are very important
for the amplitude of the correlation function, especially on small scales. We
show in Fig. \ref{fig:test} that the difference between two extreme cases:
where no stars are formed after accretion and where galaxies suffer no
stripping of stars, result in differences in the amplitude of the two-point
correlation function of $\pm20\;\%$ on large scales and almost a factor of 2 on
small scales.

\end{itemize}

We note that, although the box size of EAGLE (100 Mpc) is among the largest for
simulations of its type, it is not large enough to ensure converged clustering
properties. The lack of long wavemodes produces a few-percent excess of halos
with $M\lesssim 10^{14}\Mass$ and a larger deficiency of more massive halos.
We expect this to reduce the satellite fraction, which may affect the shape and
amplitude of overall correlation function, and might thus make our assessment of
SHAM slightly too optimistic.

Overall, our results confirm the usefulness of SHAM for interpreting and
modelling galaxy clustering. However, they also highlight the limits of current
SHAM implementations when an accuracy better than $\sim20\;\%$ is required.
Beyond this point, details of galaxy formation physics become important. For
instance, SHAM assumes that the relation between $\Vrelax$ and $\Mstar$ is
independent of the host halo mass. However, the validity of this assumption
depends on how efficiently the gas content of satellite galaxies is depleted
after accretion, on the importance of the stripping of stars in different
environments, and on the relation between $M_{DM}$ and $\Mstar$ for centrals.
EAGLE suggests that these effects depend on the host halo mass (and thus
possibly on cosmological parameters), which would break the family of
one-parameter SHAM models. 

Fortunately, it seems possible that these physical processes can be modelled,
and marginalised over, within SHAM. An interesting line of development would be
the extension of SHAM to a two-parameter model, for instance a function of
$\Vrelax$ and $\Mhalo$.  This would not only reduce the systematic biases in
the correlation function, but would also increase the predictive power of SHAM
for centrals. We plan to explore this in the future.

Naturally, as hydrodynamical simulations improve their realism, it should be
possible to better model the evolution of galaxies hosted by massive clusters,
which will lead to more accurate SHAM implementations and a more accurate
assessment of its performance. Ultimately, these developments will enable quick
and precise predictions for the clustering of galaxies in the highly non-linear
regime. In principle, this could be extended as a function of cosmology
employing, e.g., cosmology-scaling methods \citep{Angulo2010, Angulo2015}. This
opens up many interesting possibilities, such as the direct use of SHAM to
optimally exploit the overwhelmingly rich and accurate clustering measurements
that are expected to arrive over the next decade.

\section*{Acknowledgements}

We would like to thank Oliver Hahn and Peter Behroozi for useful discussions.
Most of the parameters for EAGLE galaxies are available from the database
(McAlpine+) or through interaction with the authors. This research was supported
in part by the European Research Council under the European Union's Seventh
Framework Programme (FP7/2007-2013) / ERC Grant agreement
278594-GasAroundGalaxies, GA 267291 Cosmiway, and 321334 dustygal, the
Interuniversity Attraction Poles Programme initiated by the Belgian Science
Policy Office ([AP P7/08 CHARM]). This work used the DiRAC Data Centric system
at Durham University, operated by the Institute for Computational Cosmology on
behalf of the STFC DiRAC HPC Facility (www.dirac.ac.uk). This equipment was
funded by BIS National E-infrastructure capital grant ST/K00042X/1, STFC capital
grant ST/H008519/1, and STFC DiRAC Operations grant ST/K003267/1 and Durham
University. DiRAC is part of the National E-Infrastructure. RAC is a Royal
Society University Research Fellow. J.C.M acknowledges support from the
“Fundaci\'on Bancaria Ibercaja” for developing this research. 

%----------------------------------------------
\bibliographystyle{mn2e}
\bibliography{database}
%---------------------------------------------------------------------

\begin{figure}
\begin{center}
\includegraphics[width=0.45\textwidth]{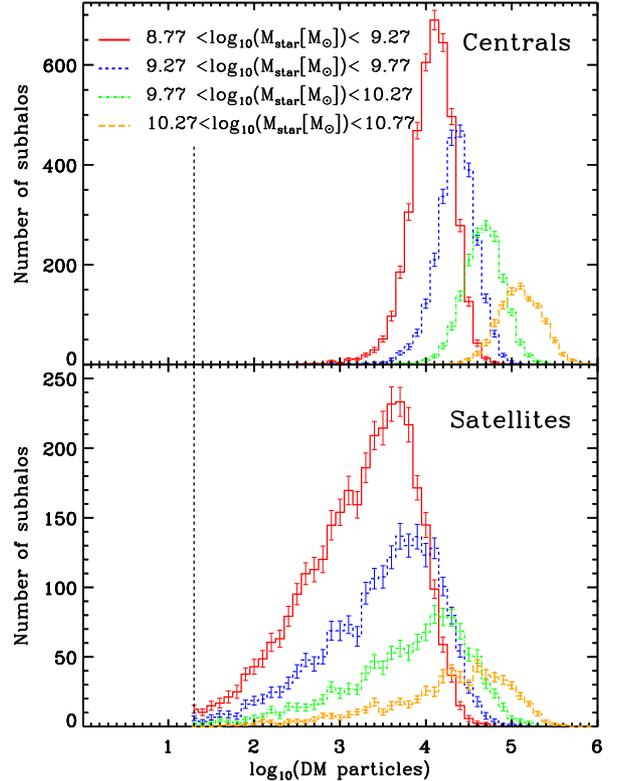}
\end{center}
\caption{
\label{fig:Resolution} Number of DM particles in subhaloes of a given $\Mstar$.
The coloured lines represent the mean PDFs of 100 realizations using $\Vrelax$
for different stellar mass bins and the errors are the standard deviation of
the 100 realizations. The top (bottom) panel shows the PDFs of centrals
(satellites). The black dashed line indicates the detection threshold of our
SUBFIND catalogues. The centrals are always resolved with more than 1000 particles.
However, the satellites have a tail in their distribution which reaches the
detection threshold.}
\end{figure}

\begin{figure*}
\begin{center}
\includegraphics[width=0.9\textwidth]{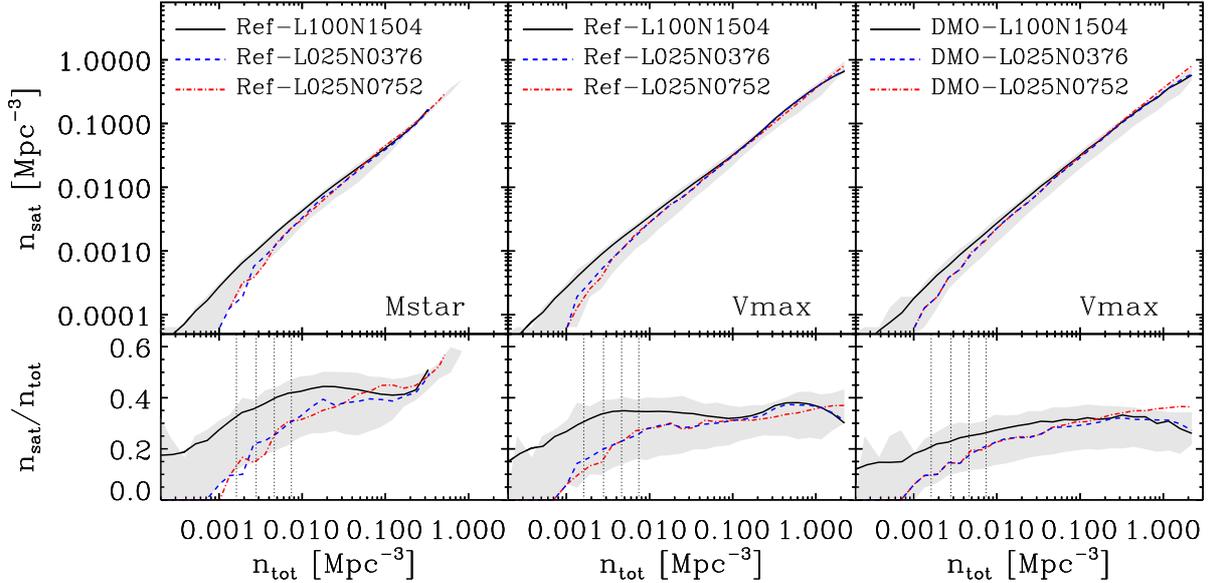}
\end{center}
\caption{
\label{fig:Res2} Number density of satellites (top panels) and satellite
fraction (bottom panels) vs. total number density. In the left, centre, and
right panels subhaloes are ordered according to $M_{\rm star}^{\rm Ref}$,
$V_{\rm max}^{\rm Ref}$, and $V_{\rm max}^{\rm DMO}$ respectively. Coloured
lines show the results for different simulations. The grey shaded areas enclose
the $68\;\%$ of the results after dividing the simulation with the largest volume
into 64 smaller boxes of $25\;\Mpc$ on a side. The regions enclosed by dotted
lines indicate the bins employed in \S\ref{sec:results}. }
\end{figure*}

\section*{Appendix A: Resolution}
\label{sec:resolution}

In this section we will present two tests that suggest that our results are not
affected by the finite mass and force resolution of the EAGLE and DMO
simulations. Specifically, we will explore the number of DM particles of the
SHAM galaxies and compare simulations with different resolutions.

In Fig. \ref{fig:Resolution} we show the PDF of the number of DM particles
associated with central (top panel) and satellite (bottom panel) SHAM galaxies.
Coloured lines show the results for different $\Mstar$ bins using $\Vrelax$.
The detection threshold of our SUBFIND catalogues (20 particles) is marked by a
vertical dashed line. The top panel shows that nearly all the central subhaloes
are resolved more than $1000$ DM particles. Satellites, on the other hand, are
resolved with fewer particles because some of them will be lost to tidal
stripping. However, since the value of $\Vrelax$ will be acquired before the
stripping begins, we do not expect this to affect our results. The only effect
that might be important is that a subhalo can fall below the detection
threshold while its counterpart galaxy is still resolved. We see that this
might be the case for a very small fraction subhaloes in the lowest $\Mstar$
bin. We quantify these effects next.

In Fig. \ref{fig:Res2} we show the number density of satellites (top panel)
and the satellite fraction (bottom panel) for three different EAGLE simulations
and their DMO counterparts. The black lines show the results for the same
simulation used in this paper (Ref--L100N1504), the blue lines for a simulation
with $25\,\Mpc$ on a side and the same resolution as Ref--L100N1504
(Ref--L025N376), and the red lines for a simulation with $25\,\Mpc$ on a side
and eight times higher mass resolution than Ref--L100N1504 (Ref--L025N752). To
estimate the cosmic variance, we divide Ref--L100N1504 into $64$ boxes
of $25\,\Mpc$ on a side; the grey shaded areas enclose the $68\;\%$ of
these boxes. The regions enclosed by vertical dotted lines in the bottom panels
indicate the bins employed in \S\ref{sec:results}.

The left two panels show that galaxies according to $\Mstar$ or $\Vmax$ produce
almost identical satellite fractions in both (Ref--L025N752) and
(Ref--L025N386), despite the former having 8 times better mass resolution. The
satellite fraction coincides with our main EAGLE run for high number densities,
but under-predicts the satellite fraction at low number densities. This,
however, is plausibly explained by cosmic variance and the lack of long wave
modes due to the smaller volume (64 times). The rightmost panel shows the DMO
versions, for which the agreement between different resolutions is even better.
Thus, this suggests that the results presented in this paper are not affected
by the numerical resolution of our simulations.

\section*{Appendix B: Correlation function calculation}

The two point correlation function (2PCF) counts the number of pairs at
different distances in relation to the number of pairs that one would have
expected from a random distribution \citep[see,
e.g.,][]{1985ApJ...292..371D,2001ASPC..252..201P}:

\begin{equation}
dP=n^2 [1+\xi(\mathbf{r}_{12})] dV_1 dV_2,
\end{equation}

\noindent where $n$ is the mean density and $\xi(\mathbf{r}_{12})$ the
correlation function. This equation describes the excess probability, compared
with a random sample, of finding a point in an element of volume $dV_2$ at a
distance $\mathbf{r}_{12}$ from another point in $dV_1$. The 2PCF is also the
Fourier transform of the power spectrum $P(\mathbf{k})$:

\begin{equation}
\xi(\mathbf{r})=\frac{1}{(2\pi)^3}\int dk^3 P(\mathbf{k}) e^{i \mathbf{k}\cdot \mathbf{x}},
\end{equation}

\noindent and the power spectrum is defined as:

\begin{equation}
\label{eq:11}
\left<\hat{\delta}(\mathbf{k})\hat{\delta}(\mathbf{k}')\right>=(2\pi)^3 \delta_{\rm D}(\mathbf{k}-\mathbf{k}')P(\mathbf{k}),
\end{equation}

\noindent where $\hat{\delta}(\mathbf{k})$ is the Fourier transform of the
density contrast and $\delta_D(\mathbf{k})$ is the Dirac delta function. We can
use this property to quickly compute the 2PCF using the fast Fourier transform
(FFT). To calculate the 2PCF, we follow the following steps:

\begin{itemize}

\item We divide the simulation cube into $1024^3$ boxes of 97.6 $\kpc$ on a
side. We determine in each box the density contrast using a cloud-in-cell (CIC)
scheme. The density contrast is defined as:

\begin{equation}
\delta(\mathbf{x})=\frac{N-\left<N\right>}{\left<N\right>},
\end{equation}

\noindent where $N$ is the number of subhaloes inside one box and
$\left<N\right>$ is the total number of subhaloes in the simulation cube.

\item The Fourier transform (FT) of the density field is:
\begin{equation}
\hat{\delta}(\mathbf{k})=\int dx^3e^{-i \mathbf{k}\cdot \mathbf{x}}\delta(\mathbf{x}),
\end{equation}
we compute this FT using version 3.3.3 of the Fastest Fourier Transform in the
West ({\small FFTW3}; http://www.fftw.org/), a compilation of C routines for
computing the discrete Fourier transform.

\item We calculate $P(\mathbf{k})$ using equation \ref{eq:11} and then we
subtract the Poisson noise. The Poisson noise arises from sampling a continuous
distribution with a discrete number of objects. It scales as $1/n$, where $n$ is
the number density of objects.

\item The next step is to go back to real space by computing the FT of
$P(\mathbf{k})$, yielding the 2PCF.

\item Finally, we spherically average the correlation function obtaining the 3D
2PCF $\xi(|\mathbf{r}|)$.
\end{itemize}

By dividing the simulation cube into different number of cells, we verified that
using $1024^3$ boxes represents the clustering beyond $0.3\;\Mpc$ faithfully.

\label{lastpage} \end{document}